\documentclass[aps,preprint,onecolumn,12pt,nofootinbib]{revtex4}
\usepackage[utf8]{inputenc}
\usepackage[english]{babel}

\usepackage{graphicx}
\usepackage{subcaption}
\usepackage{caption}
\usepackage{tikz}
\usetikzlibrary{quotes,angles,arrows,decorations.markings}

\usepackage{amsmath,amssymb,amsfonts,mathrsfs}
\usepackage{bm,bbm,dsfont,braket,tensor,esint}
\usepackage{cancel,booktabs,multirow,array,makecell}
\newcolumntype{C}{>{\centering\arraybackslash}p{1.8cm}}

\usepackage{color,xcolor}
\usepackage{float}
\usepackage{lipsum}
\usepackage{enumerate}
\usepackage{ulem}
\newif\ifshowchanges
\showchangesfalse   

\ifshowchanges
    \usepackage{changes}
    \setaddedmarkup{\textcolor{red}{#1}}
    \setdeletedmarkup{\textcolor{gray}{\sout{#1}}}
\else
    \usepackage[final]{changes}
\fi

\ifshowchanges
    \newcommand{\revision}[1]{\textcolor{red}{#1}}
    \newcommand{\replace}[2]{\textcolor{gray}{\sout{#1}}\textcolor{red}{#2}}
    \newcommand{\redsout}[1]{\textcolor{red}{\sout{\textcolor{black}{#1}}}}
    
\else
    \newcommand{\revision}[1]{#1}
    \newcommand{\replace}[2]{#2}
    \newcommand{\redsout}[1]{}
\fi

\usepackage{hyperref}
\hypersetup{
    colorlinks,
    breaklinks,
    citecolor=blue,
    filecolor=green,
    linkcolor=purple,
    urlcolor=red,
    citecolor=[rgb]{0,0.0,1.0},
    urlcolor=[rgb]{0.0,0.0,1.0},
    linkcolor=[rgb]{0,0.5,0.9}
}

\usepackage{slashed}

\definecolor{lime}{HTML}{A6CE39}
\DeclareRobustCommand{\orcidicon}{%
    \begin{tikzpicture}
    \draw[lime, fill=lime] (0,0) 
    circle [radius=0.16] 
    node[white] {{\fontfamily{qag}\selectfont \tiny ID}};
    \draw[white, fill=white] (-0.0625,0.095) 
    circle [radius=0.007];
    \end{tikzpicture}
    \hspace{-2mm}
}

\foreach \x in {A, ..., Z}{%
    \expandafter\xdef\csname orcid\x\endcsname{\noexpand\href{https://orcid.org/\csname orcidauthor\x\endcsname}{\noexpand\orcidicon}}
}

\begin{document}

\title{Probing Hernquist dark matter \replace{with black hole shadows}{through the optical appearance of black holes}: A comprehensive study of various accretions}

\author{Yuxuan Shi\orcidA{}}
\email{shiyx2280771974@gmail.com}
\affiliation{School of Physics, East China University of Science and Technology, Shanghai 200237, China}

\author{Hongbo Cheng}
\email{hbcheng@ecust.edu.cn}
\affiliation{School of Physics, East China University of Science and Technology, Shanghai 200237, China}
\affiliation{The Shanghai Key Laboratory of Astrophysics, Shanghai, 200234, China}

\begin{abstract}
\replace{The shadow of a black hole is critically dependent on the surrounding accreting matter.}{The observational appearance of a black hole is critically dependent on the surrounding accreting matter, in particular on the central brightness depression and photon ring structure.} We perform a systematic comparative analysis of the observational signatures of a Schwarzschild black hole embedded in a Hernquist dark matter (DM) halo under three distinct accretion scenarios: a geometrically thin disk, a static spherical flow, and an infalling spherical flow. For the thin disk model, we find that direct emission dominates the total observed intensity, while the size and brightness of the lensing and photon rings serve as sensitive probes of the Hernquist DM parameters. From a geometric perspective, the Hernquist DM halo significantly enlarges the photon sphere, resulting in an observable \replace{shadow radius}{critical curve radius} approximately $2\%$ to $30\%$ larger than in the vacuum case. Regarding the radiative signatures, the measured intensity profiles, which rely on the particular accretion models, show a general brightness suppression, which is especially affected by the Doppler de-boosting in the infalling scenario. Our results suggest that \replace{the size and brightness profile of a black hole shadow provide}{the size of the central brightness depression and the brightness profile of the black hole image provide} a valuable theoretical framework for constraining the distribution of dark matter in galactic centers.
\end{abstract}

\maketitle


\section{Introduction}
The observation of black hole shadows has entered a new era, establishing them as a ground-breaking tool for probing strong gravity regimes. The fundamental nature of gravity has been further illuminated by analysing the critical behaviour of photons close to black holes. Recent advances in the Event Horizon Telescope's (EHT) imaging of the supermassive black holes $\text{M87}^*$~\cite{EHT1,EHT2,EHT3,EHT4,EHT5,EHT6,EHT7,EHT8,EHT9} and $\text{SgrA}^*$~\cite{ghez2005stellar,boehle2016improved,sgr1,sgr2,sgr3,sgr4,sgr5,sgr6,sgr7,sgr8} have shown shadow features that not only support general relativity (GR)~\cite{hawking1972black,bardeen1973four,wald1994quantum,wald2010general} predictions but also provide a novel window for probing black holes. The black hole shadow, also known as the photon capture zone, is the central dark area observed in these images~\cite{perlick2022calculating}. It is encircled by a compact, asymmetrical and bright circular structure. This phenomenon, known as the photon ring, results from light rays from infinity being deflected by the strong gravitational field as they approach the black hole. 
In backward ray-tracing methods, light rays that approach the critical impact parameter asymptotically converge to the bound photon orbit. The so-called photon sphere, which projects the shadow of the black hole, consists of this confined photon orbit. An angular radius equation for the shadow of Schwarzschild black holes has been established by Synge and Luminet using a critical impact parameter of $b_p=3\sqrt{3}M$~\cite{synge1966escape,luminet1979image}. The first image of a black hole surrounded by a thin accretion disk was analytically produced by the authors in Ref.~\cite{luminet1979image}, displaying primary and secondary images that appear outside the black hole's shadow. Bardeen later studied the D-shaped shadow of Kerr black holes, showing that the shape of the shadow can be changed by the spin of the black hole~\cite{bardeen1973timelike,chandrasekhar1998mathematical}. Additionally, extensive research has been conducted on black hole shadows in higher-dimensional spacetime and other modified gravity theories~\cite{amarilla2010null,amarilla2012shadow,amir2018shadows,mizuno2018current,eiroa2018shadow,banerjee2020silhouette,bambi2019testing,vagnozzi2023horizon,khodadi2024event}. In contrast to single shadows, studies have also suggested multiple shadows for wormholes and black holes~\cite{huang2016double,guerrero2021double}. While the findings of the EHT mainly relate to black holes in GR, they provide significant opportunities to investigate other compact objects in theories other than GR.

To investigate the observational characteristics of black hole images and search for potential indications of novel physics, it is essential to employ simplified accretion models. Schwarzschild black holes were studied for the spherically symmetric accretion model, another accretion scenario, and it was shown that under such accretion settings the shadow has robust properties, where its shape and size are determined by spacetime geometry rather than accretion details~\cite{gralla2019black,tsukamoto2014constraining,tsukamoto2018black}. Conversely, investigations of optically thin and geometrically thin accretion models show that the appearance of the \replace{shadow}{central brightness depression} is strongly related to the position of the emission source~\cite{zeng2020influence,zeng2020shadows,li2021shadows,he2022shadow,peng2021influence,zeng2022shadows,zeng2025holographic}. Direct emission, lensing rings, and photon rings are the three main components that these models usually appear as in the seen image. These distinct signatures offer a potential means to distinguish GR black holes from other compact objects or black holes in modified gravity theories via observational data~\cite{zeng2020influence,zeng2020shadows,li2021shadows,he2022shadow,peng2021influence,zeng2022shadows,zeng2025holographic}. Consequently, EHT observations of black hole shadows are crucial for constraining parameters in modified gravity theories and testing fundamental physics~\cite{shaikh2019shadows,davoudiasl2019ultralight,roy2020evolution}.

The environment surrounding the black hole, in particular the distribution of DM, offers a compelling explanation for possible departures from conventional predictions, in addition to modified gravity theories. The discovery of massive elliptical and spiral galaxies marked a significant development in the search for DM~\cite{rubin1980rotational}. Succeeding studies indicated that about $90\%$ of a galaxy's mass is composed of DM~\cite{persic1996universal}. Estimating DM's contribution close to the galactic centre is especially important since there is strong evidence that DM halos around astrophysical black holes~\cite{EHT1,EHT6,sofue2013mass,boshkayev2019model,jusufi2019black,konoplya2019shadow,konoplya2022solutions,hou2018black,liang2023thermodynamics,gohain2024thermodynamics,ovgun2025black,jha2025thermodynamics}. Several DM distributions are covered by the Dehnen density distribution, which is frequently used to describe dwarf galaxies~\cite{dehnen1993family,mo2010galaxy}. These investigations usually use the $\text{Dehnen}-(1,4,1)$ type, corresponding to the Hernquist DM distribution, is widely used to describe elliptical galaxies. The Hernquist profile provides a mathematically tractable description that is consistent with the dynamical evidence of galaxies. Nevertheless, more detailed examination is necessary to understand the distinct observational signatures  of Schwarzschild black holes embedded in Hernquist DM halos, especially under realistic accretion scenarios. Although the effects of Hernquist DM halos on the photon sphere radius and shadow radius have been discussed in Ref.~\cite{jha2025thermodynamics,ahmed2025observable,feng2026shadow}, accretion models remains less explored. Furthermore, quantitative studies linking halo parameters, such as central density and core radius, to the structure of lensing and photon rings are limited, as most existing works concentrate on vacuum or uniform DM backgrounds~\cite{gralla2019black,zeng2020influence,zeng2020shadows,li2021shadows,he2022shadow,peng2021influence,zeng2022shadows,zeng2025holographic,shaikh2019shadows,shi2024shadow,feng2026shadow,amir2018shadows,eiroa2018shadow,amarilla2012shadow,konoplya2019shadow,mizuno2018current,meng2023images}. This work aims to bridge this gap by conducting a systematic comparison of accretion models within a Hernquist DM halo. Addressing this gap is essential for using black hole shadows as accurate probes to restrict DM characteristics.

In this paper, we present a systematic comparative study of the shadow of Schwarzschild-Hernquist black holes with different accretion models. In order to clarify the associated impacts of Hernquist DM halo parameters on the photon sphere and observed brightness, we thoroughly investigated three thin disk accretion models and two spherical accretion models. Our results demonstrate that different accretion regimes produce \replace{different shadow images}{distinctly different images} under the same background geometry. Specifically, pictures made up of the \replace{shadow}{central brightness depression}, lensing ring, and photon ring are created via thin disk accretion. The shadow and photon ring are spherically symmetric for spherical accretion. A darker shadow than under static conditions results from the extra Doppler effect introduced by falling accretion~\cite{li2021observational,li2021shadows,zeng2020influence,zeng2020shadows,zeng2022shadows,zeng2025holographic}. We aim to close a gap in the references on Hernquist metrics and accretion models by revealing scaling rules for shadow size and brightness as functions of the Hernquist DM halo distribution. This offers a potential diagnostic framework for constraining Hernquist DM characteristics.

The structure of this paper is as follows: Sec.\ref{sec2} derives the null geodesic under the Hernquist metric, defining the photon sphere radius and critical impact; Sec.\ref{sec3} analyses the thin disk accretion model, including the contribution regions of direct radiation, lensing rings, and photon rings, and relates the emission intensity to the observed appearance; Sec.\ref{sec4} examines static and infalling spherical accretion models, contrasting how varying Hernquist DM halo parameters affect shadow brightness; finally, Sec.\ref{conclusion} summarizes the findings, emphasising the potential interpretative power of Hernquist DM halos for shadows and outlining their prospects for detecting DM distributions in galactic centres.

\revision{Before proceeding, we clarify the terminology adopted in this work. The critical curve refers to the projection of the photon sphere onto the observer's image plane, characterized by the critical impact parameter $b_p$. The term shadow, in its classical sense following Falcke~\cite{falcke2000viewing}, denotes the region enclosed by the critical curve from which no photons reach the distant observer. However, as emphasized by Gralla~\cite{gralla2019black}, the observed central brightness depression does not necessarily coincide with the critical curve and depends on the properties of the accretion flow~\cite{falcke2000viewing,gralla2019black,macedo2024optical,feng2026shadow}. For the spherical accretion models, the boundary of the central dark area closely traces the critical curve, so the classical notion of the shadow applies directly. In contrast, for the thin disk models, the observed dark region is shaped by the emission profile and may differ from the classical shadow.}

\section{Light deflection and photon orbit}
\label{sec2}
In this work, we model a Schwarzschild black hole immersed in a DM halo that is described by the Hernquist density profile~\cite{dehnen1993family,mo2010galaxy},
\begin{align}
\rho(r)=\rho_c\left(\dfrac{r}{r_s}\right)^{-1}\left[1+\dfrac{r}{r_s}\right]^{-3},
\end{align}
where $\rho_c$ and $r_s$ represent the center density and core radius of the DM halo, respectively. The Hernquist DM mass profile can be expressed as follows,
\begin{align}
\label{DM_mass}
M_H = \int_0^r4\pi\rho(r')r'^2\mathrm{d}r'=\dfrac{2\pi r^2r_s^3\rho_s}{(r+r_s)^2}.
\end{align}
According to the derivation in Ref.~\cite{jha2025thermodynamics}, the spherically symmetric metric for the combined system of black hole and Hernquist DM halo may be assumed,
\begin{align}
\label{metric_fun}
\mathrm{d}s^2=-f(r)\mathrm{d}t^2+\dfrac{\mathrm{d}r^2}{f(r)}+r^2\left(\mathrm{d}\theta^2+\sin^2\theta\mathrm{d}\varphi^2\right),
\end{align}
where the metric function is the linear superposition of the two,
\begin{align}
f(r)=f_{\text{H}}(r)+f_{\text{BH}}(r).
\end{align}
The tangential velocity of test particles $v_t$ and the mass profile $M_{\text{H}}$ in the halo determine the redshift function $f_{\text{H}}(r)$ via the relation~\cite{jha2025thermodynamics,ahmed2025observable},
\begin{align}
\label{relation}
v_t^2=\dfrac{M_{\text{H}}}{r}=r\dfrac{\mathrm{d}}{\mathrm{d}r}\ln\sqrt{f_{\text{H}}(r)}.
\end{align}
The redshift function $f_{\text{H}}(r)$ takes on an exponential shape when this relation is integrated. Assuming the halo potential is modest in relation \eqref{relation} to the central black hole term, a linear approximation is made to the halo contribution to generate the analytically tractable form,
\begin{align}
\label{fH}
f_{\text{H}}(r)&=\exp\left(-\dfrac{4\pi\rho_sr_s^3}{r+r_s}\right)\notag\\
&\simeq1-\dfrac{4\pi\rho_sr_s^3}{r+r_s}.
\end{align}
Importantly, even though this metric was obtained using the aforementioned approximation, it is an accurate solution to the Einstein field equations in terms of its physical status~\cite{jha2025thermodynamics},
\begin{align}
R_{\mu\nu}-\dfrac{1}{2}Rg_{\mu\nu}
=\kappa^2\left(T_{\mu\nu}^\text{H}+T_{\mu\nu}^\text{B}\right),
\end{align}
The energy-momentum tensor corresponds to an anisotropic fluid for the redshift function \eqref{fH} in the line element \eqref{metric_fun} might be expressed as
\begin{align}
T_{\mu}^{\nu}=g^{\nu\delta}T_{\mu\delta}=\text{diag}\left[-\rho,p_r,p,p\right].
\end{align}
The Schwarzschild black hole's energy-momentum tensor $T_{\mu\nu}^B = 0$ since it is a vacuum solution. As a result, the dark matter tensor $T_{\mu\nu}^{\text{H}}$ dominates the source term of the complex system. This means that $f_{\text{BH}}(r)$ must be reduced to the Schwarzschild solution in the vacuum in order for the equations to be solved~\cite{jha2025thermodynamics,ahmed2025observable},
\begin{align}
\label{fBH}
f_{\text{BH}}(r)=-\dfrac{2M}{r}.
\end{align}

At short radii $(r\ll r_s)$, this distribution shows an isothermal feature of $\rho\propto r^{-1}$, but at high radii $(r\gg r_s)$, it decays to $\rho\propto r^{-4}$\cite{jha2025thermodynamics}. For details of the computational process, we refer the reader to Refs.~\cite{mo2010galaxy,yang2024black,xu2018black}. Here, the metric of a Schwarzschild black hole immersed in the Hernquist DM halo surroundings can be written as follows~\cite{jha2025thermodynamics,ahmed2025observable,feng2026shadow}:
\begin{align}
\label{metric}
f(r)=1-\dfrac{2M}{r}-\dfrac{4\pi\rho_cr_s^3}{r+r_s}.
\end{align}

\begin{figure}[t]
\centering
\includegraphics[width=0.8\textwidth]{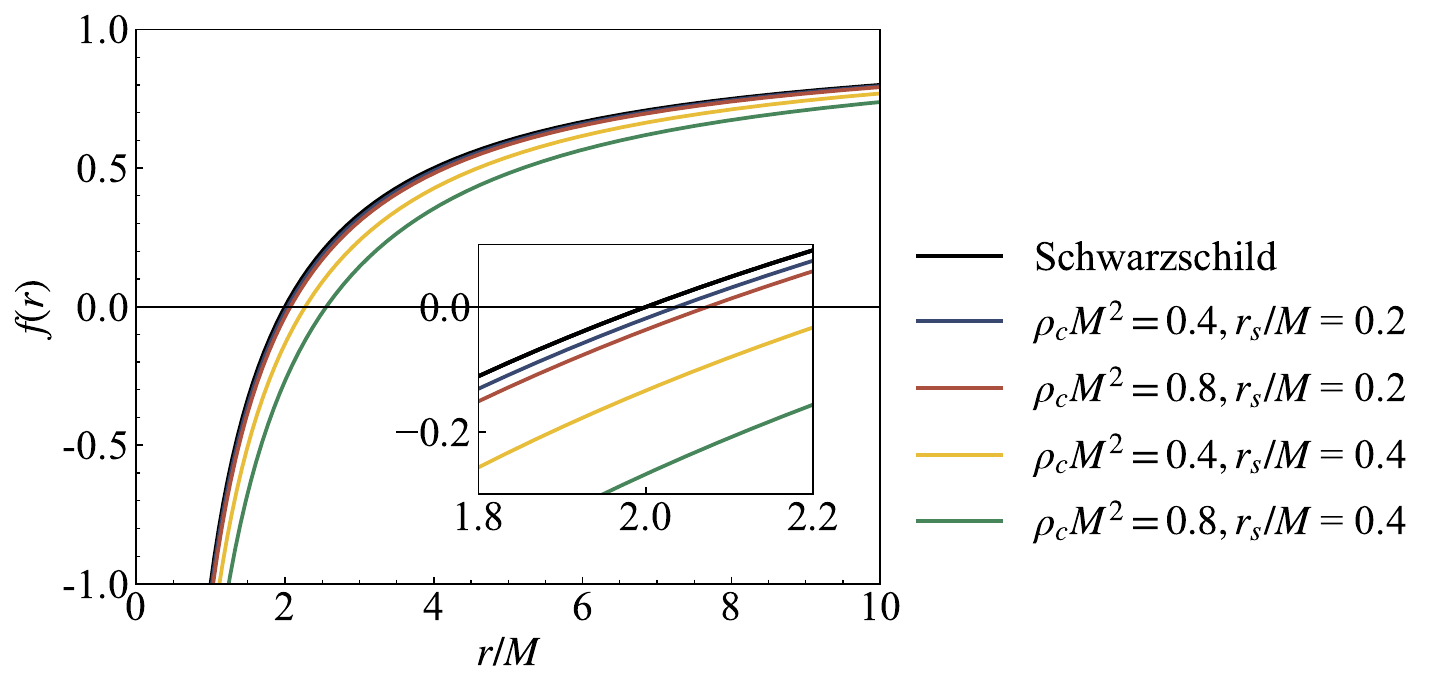}
\caption{The metric function $f(r)$ as a function of the radial coordinate $r/M$ for different Hernquist DM halo parameters. The horizontal line represents $f(r)=0$. A close-up of the area close to the event horizon is shown in the inset. In contrast to the Schwarzschild case (black line), it is clear that the zero-crossing point, which determines the event horizon radius $r_h$, spreads outward as the halo parameters $\rho_c$ and $r_s$ grow.}
\label{fig:fr}
\end{figure}

Here, the Newton constant $G$ has been set to $1$ for simplicity. Naturally, this black hole metric \eqref{metric} reduces to the Schwarzschild black hole in the limit $\rho_c\to0$ or $r_s\to0$. 
In Fig.\ref{fig:fr}, we depict the behavior of the metric function $f(r)$ to show how the Hernquist DM halo affects spacetime. The largest root of the equation $f(r_h)=0$ defines the event horizon radius $r_h$. This critical point in the Schwarzschild scenario is located at $r_h=2M$. However, the DM halo causes the curve to cross the zero axis at a wider radius, as seen in Fig.\ref{fig:fr}. In particular, a monotonic expansion of the event horizon is induced by an increase in either the core radius $r_s$ or the central density $\rho_c$. This geometric change indicates that the input from the DM halo may successfully boost the black hole's gravitational capture ability.

\revision{The Hernquist profile adopted here belongs to a broader family of dark matter density models, each targeting distinct astrophysical scenarios. The Navarro--Frenk--White (NFW)  profile~\cite{navarro1996structure,navarro1997universal}, arising from $\Lambda$CDM $N$-body simulations, provides a near-universal description of cold dark matter halos but suffers from a logarithmically divergent total mass. The Burkert profile~\cite{burkert1995structure,salucci2000dark}, featuring a constant-density core, was introduced to accommodate the flat rotation curves of dwarf and low-surface-brightness galaxies. The Einasto profile~\cite{einasto1965construction,merritt2006empirical} offers a flexible three-parameter fit that is statistically superior to NFW in high-resolution simulations. Within the Dehnen family~\cite{dehnen1993family}, characterized by an inner-slope index $\gamma\in[0,3)$, the Hernquist model corresponds to the $\gamma=1$ case~\cite{hernquist1990analytical}. This choice is particularly well suited for studying black holes embedded in elliptical galaxies: its density--potential pair is expressible in elementary functions, allowing the metric to be obtained analytically~\cite{konoplya2022solutions,macedo2024optical,feng2026shadow}; its total mass converges exactly, eliminating the truncation ambiguity of NFW-based models; and the $r^{-1}$ inner cusp combined with the $r^{-4}$ outer falloff faithfully reproduces the de~Vaucouleurs $R^{1/4}$ surface-brightness law characteristic of elliptical galaxies~\cite{hernquist1990analytical,de1948recherches}.}

Our goal is to investigate the Schwarzschild-Hernquist black hole's null geodesic. We restrict our analysis to the equatorial plane, $\theta=\pi/2$ and $\dot{\theta}=0$~\cite{synge1966escape,bardeen1972rotating,gralla2020lensing}. Therefore, the Lagrangian $\mathcal{L}$ for a particle in this spacetime can be easily obtained,
\begin{align}
\label{Lagranian}
\mathcal{L}
&=\frac{1}{2}g_{\mu\nu}\dot{x}^{\mu}\dot{x}^{\nu} \notag\\
&=\frac{1}{2}\left[-f(r)\dot{t}^{2}+\dfrac{\dot{r}^{2}}{f(r)}+r^2\dot{\varphi}^{2}\right],
\end{align}
where, $\dot{x}^{\mu}=\partial x^{\mu}/\partial\chi$ is the four-velocity of the photon and $\chi$ is the affine parameter. Since the coefficient of the metric equation cannot be directly found using the $t$ and $\theta$ coordinates, energy and angular momentum are represented by the conserved variables $E$ and $L$, respectively. That's~\cite{shaikh2019shadows,he2022shadow,li2021shadows,shi2024shadow,feng2026shadow}
\begin{align}
\label{energy}
-E&=\dfrac{\partial\mathcal{L}}{\partial\dot{t}}=-f(r)\dot{t},\\
\label{momentum}
L&=\dfrac{\partial\mathcal{L}}{\partial\dot{\varphi}}=r^{2}\dot{\varphi}.
\end{align}
By taking into account that $g_{\mu\nu}\dot{x}^{\mu}\dot{x}^{\nu}=0$ for the null geodesic, and then using Eqs.\eqref{energy}-\eqref{momentum} to solve Eq.\eqref{Lagranian}, it produces~\cite{amir2018shadows,eiroa2018shadow,shaikh2019shadows,he2022shadow,li2021shadows}
\begin{align}
\label{dot_t}
\dot{t}&=\dfrac{1}{b}\left(1-\dfrac{2M}{r}-\dfrac{4\pi\rho_cr_s^3}{r+r_s}\right)^{-1},\\
\label{dot_phi}
\dot{\varphi}&=\pm\dfrac{1}{r^{2}},\\
\label{dot_r}
\dot{r}^{2}&=\dfrac{1}{b^{2}}-\dfrac{1}{r^{2}}\left(1-\dfrac{2M}{r}-\dfrac{4\pi\rho_cr_s^3}{r+r_s}\right).
\end{align}
In this case, the impact parameter is $b=|L|/E$, and we redefine the affine parameter $\chi\to\chi/|L|$, and the sign $\pm$ denotes the light ray's clockwise and anticlockwise directions. When the effective potential $V_{\text{ph}}$ is included, we may rewrite Eq.\eqref{dot_r} as
\begin{align}
\dot{r}^2+V_{\text{ph}}=\dfrac{1}{b^2},
\end{align}
where
\begin{align}
V_{\text{ph}}=\dfrac{1}{r^{2}}\left(1-\dfrac{2M}{r}-\dfrac{4\pi\rho_cr_s^3}{r+r_s}\right).
\end{align}
The conditions for a photon sphere orbit are $\dot{r}=0$ and $\ddot{r}=0$, which translate to
\begin{align}
V_{\text{ph}}&=\dfrac{1}{b^2},\\
\dfrac{\mathrm{d}V_{\text{ph}}}{\mathrm{d}r}&=0.
\end{align}
This equation shows that when the photon sphere radius $r_p$ changes, the black hole's effective potential reaches a maximum, which corresponds to real and positive roots. Consequently, the connection between the critical impact parameter $b_p$ and the photon sphere radius $r_p$ is as follows\cite{gralla2020lensing,li2021observational,fathi2023observational,he2022shadow,shi2024shadow}:
\begin{align}
\label{sol_rp_bp_1}
r_p^2&=b_p^2f(r),\\
\label{sol_rp_bp_2}
\dfrac{r_p^3}{2}\dfrac{\mathrm{d}f(r)}{\mathrm{d}r}&=b_p^2f(r)^2.
\end{align}
The radius $r_p$ and the impact parameter $b_p$ of the photon sphere are represented in Eqs.\eqref{sol_rp_bp_1}-\eqref{sol_rp_bp_2}, respectively. Tab.\ref{tab:rh+rp+bp} displays numerical results for the event horizon radius $r_h$, photon sphere radius $r_p$, and impact parameter $b_p$ by choosing various values for $\rho_c$ and $r_s$. It is evident that as $\rho_c$ and $r_s$ increase, the event horizon radius $r_h$, the photon sphere radius $r_p$, and the impact parameter $b_p$ all rise, exceeding their values in the Schwarzschild case. This increase is a direct physical consequence of the additional mass-energy contributed by the DM halo, which enhances the spacetime curvature outside the event horizon. To maintain an unstable circular orbit within this stronger gravitational field, a photon must possess a larger angular momentum. For a distant observer, this corresponds to a larger critical impact parameter $b_p$, and the radius of the orbit itself $r_p$ is consequently shifted outwards. In essence, the DM halo enlarges the ability of the black hole for capturing photons.

\begin{table}[!tp]
\centering
\caption{The numerical results of the event horizon $r_h$, the photon sphere $r_p$ and the impact parameter $b_p$ of Schwarzschild-Hernquist black holes with $M=1$.}
\label{tab:rh+rp+bp}
\tabcolsep=0.01\linewidth
\begin{tabular}{@{}c|ccc|ccc|ccc@{}}
\toprule[1.2pt]
& $r_h/M$ & $r_p/M$ & $b_p/M$ & $r_h/M$ & $r_p/M$ & $b_p/M$ & $r_h/M$ & $r_p/M$ & $b_p/M$ \\ \cmidrule[1.2pt](l){2-4} 
\cmidrule[1.2pt](l){5-7} 
\cmidrule[1.2pt](l){8-10} 
& \multicolumn{3}{c|}{$r_s/M=0.2$} 
& \multicolumn{3}{c|}{$r_s/M=0.3$} 
& \multicolumn{3}{c}{$r_s/M=0.4$} \\
\cmidrule(lr){2-4}
\cmidrule(lr){5-7}
\cmidrule(lr){8-10}
\midrule[1.2pt]
$\rho_cM^2=0.4$ & 2.03662 & 3.05545 & 5.29421 & 2.11888 & 3.18068 & 5.51839 & 2.27357 & 3.41689 & 5.94454 \\
$\rho_cM^2=0.6$ & 2.05497 & 3.08324 & 5.34332 & 2.17894 & 3.27188 & 5.68068 & 2.41395 & 3.63042 & 6.32576 \\
$\rho_cM^2=0.8$ & 2.07335 & 3.11107 & 5.39249 & 2.23937 & 3.36361 & 5.84369 & 2.55634 & 3.84675 & 6.71082 \\
\bottomrule[1.2pt]  
\end{tabular}
\end{table}

\begin{figure}[t]
\centering
\includegraphics[width=\textwidth]{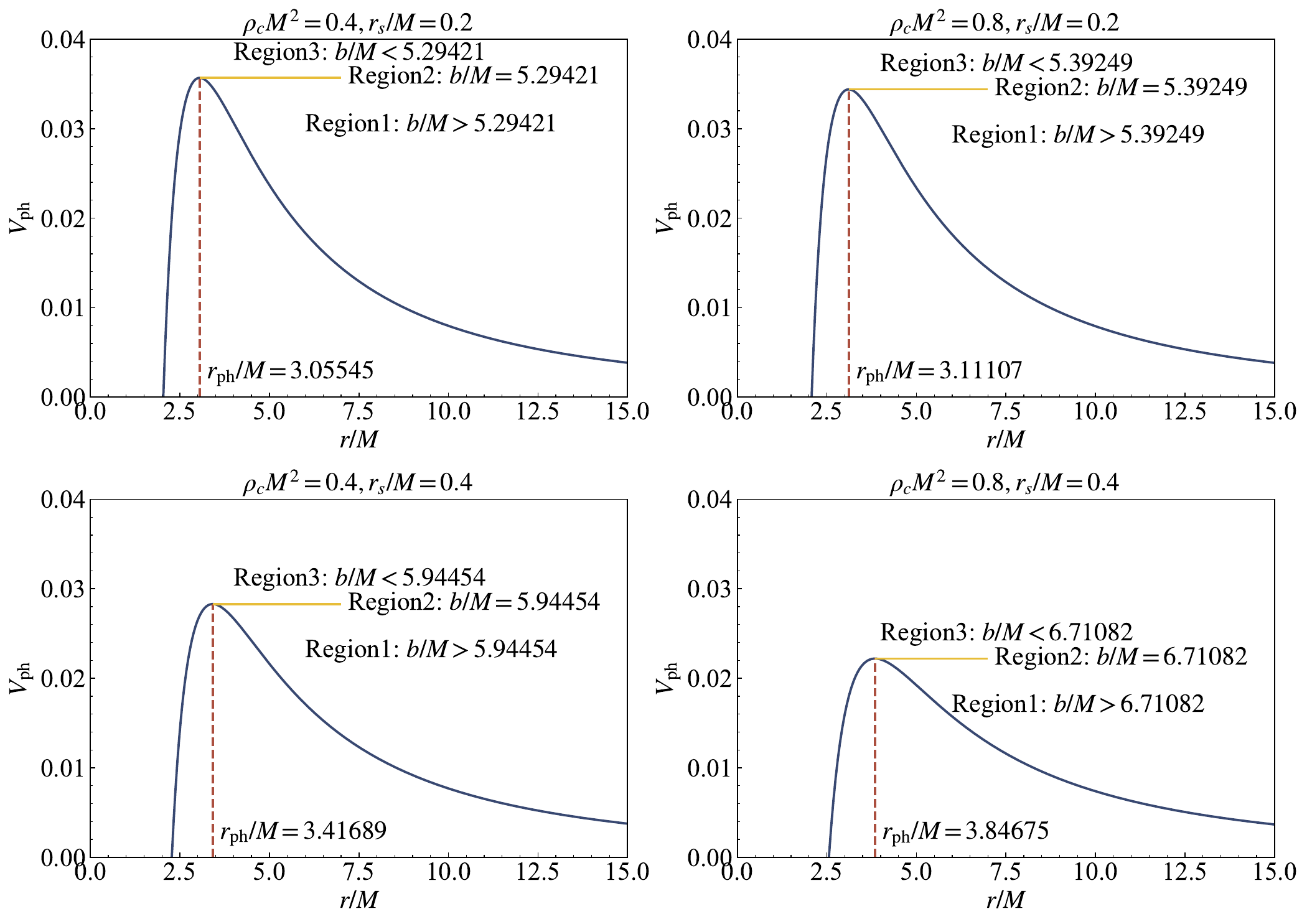}
\caption{The effective potential $V_{\text{ph}}$ as a function of the radius $r/M$ for the Schwarzschild-Hernquist black holes with varying Hernquist DM parameters. We set $M=1$ here. The peak of each curve corresponds to the photon sphere radius $r_{ph}$, which determines the critical impact parameter $b_p$. Three different physical regions of the photon trajectories are distinguished by the critical value: scattering to infinity (Region 1, $b>b_p$), critical unstable orbit (Region 2, $b=b_p$) and capture (Region 3, $b<b_p$).}
\label{fig:Vph}
\end{figure}

\revision{It is important to note that the parameter space explored in this work extends beyond the currently observationally favored region. The EHT observations of M87$^*$ constrain the angular shadow size to approximately $42\pm3\,\mathrm{\mu as}$~\cite{EHT1}, corresponding to a critical impact parameter $b_p/M_{\text{BH}}\lesssim5.56$~\cite{macedo2024optical,feng2026shadow}. From Tab.~\ref{tab:rh+rp+bp}, the configurations with $r_s/M=0.2$ (yielding $b_p/M\approx5.29$--$5.39$) remain compatible with this constraint. However, configurations with $r_s/M=0.4$ produce significantly larger values ($b_p/M\approx5.94$--$6.71$), exceeding the EHT bound. In particular, the combination $\rho_cM^2=0.8$ and $r_s/M=0.4$ yields a shadow enlargement of $\sim29\%$ relative to the Schwarzschild case, well above the current observational uncertainty of $\sim10\%$. Ref.~\cite{feng2026shadow} has derived an upper bound on the halo compactness of $\mathcal{C}\leq0.092$ from the EHT data. We emphasize that the inclusion of these large-parameter configurations is intentional: they serve as a phenomenological exploration of the theoretical upper bounds on the Hernquist DM effects, rather than as claims of astrophysical viability. A detailed assessment of the observational constraints is provided in Sec.~\ref{conclusion}.}

Fig.\ref{fig:Vph} shows the effective potential $V_{\text{ph}}$ under various $\rho_c$ and $r_s$ conditions. As an example, for case $\rho_cM^2=0.4/M^2$ and $r_s/M=0.3M$ , Fig.\ref{fig:Vph} shows that the effective potential $V_{\text{ph}}$ is zero near the event horizon and only occurs when $r\geq r_p$. It grows monotonically from this point and reaches its maximum at the photon sphere $r_p$. After then, $V_{\text{ph}}$ progressively decreases to $0$ as $r$ rises from the photon sphere towards infinity. The effective potential $V_{\text{ph}}$ naturally forms a potential barrier for incident light rays. Some light rays traveling inward from infinity are reflected by this barrier. The trajectories of these rays correspond to those in Region 1 $(b>b_p)$ in Fig.\ref{fig:Vph}. In Region 3 $(b<b_p)$ of Fig.\ref{fig:Vph}, certain rays will not pass over the barrier and will eventually fall into the black hole. Additionally, in Region 2 $(b=b_p)$, rays will asymptotically reach the photon sphere's orbit before continuing indefinitely around the black hole. 

\begin{figure}[!tp]
\centering
\includegraphics[width=\textwidth]{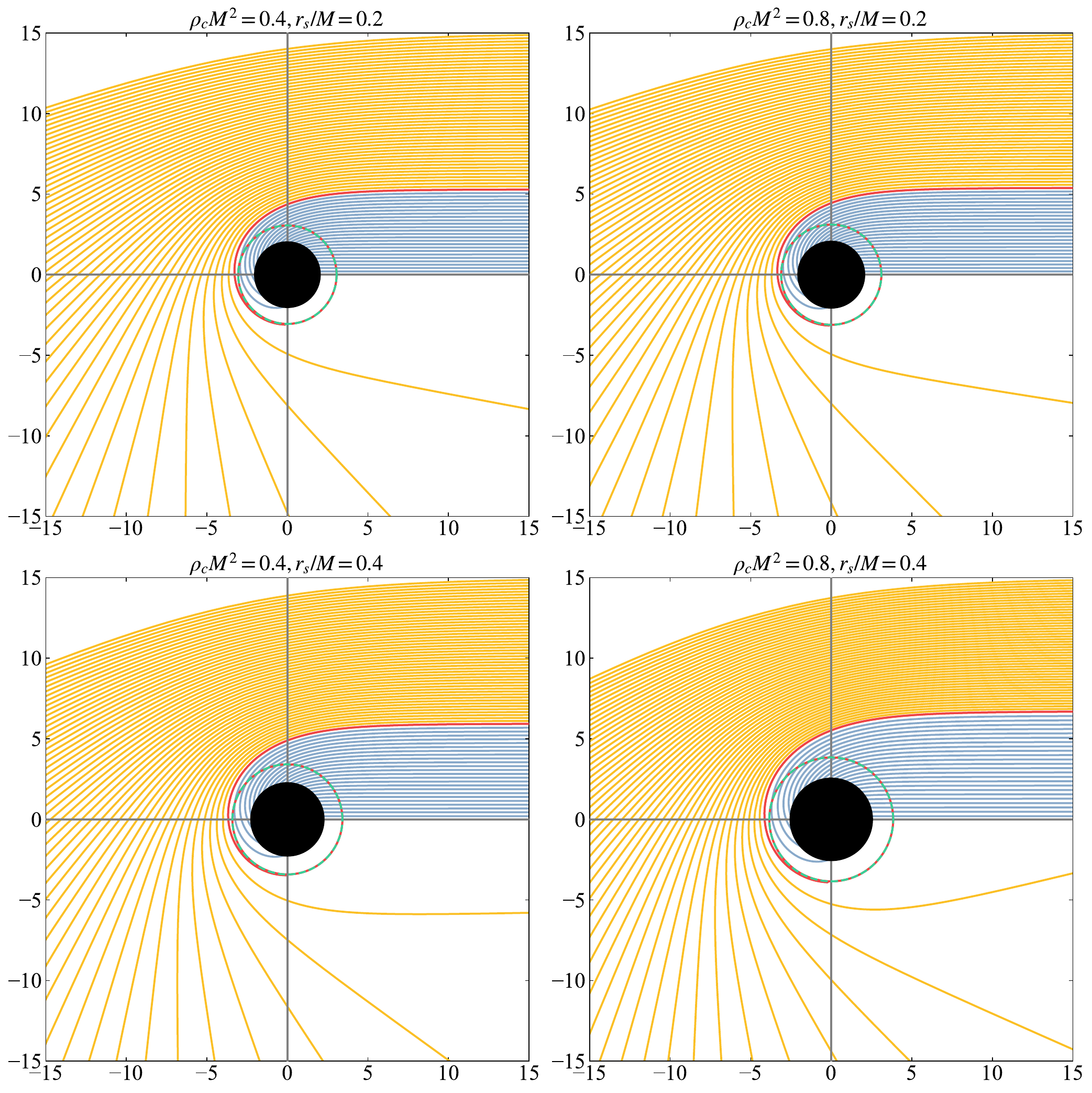}
\caption{The behaviour of light trajectories under various $\rho_c$ and $r_s$ in the polar coordinate system $(r,\varphi)$. The solid black disk and the green dashed circle indicates the event horizon and the photon sphere, respectively. The yellow lines correspond to scattered rays $(b>b_p)$, the blue lines to captured light $(b<b_p)$ and the red one means the critical unstable orbit $(b=b_p)$.}
\label{fig:trajectory_1}
\end{figure}

In order to determine the trajectory of light rays, the equation of motion for photons is reformulated using Eqs.\eqref{dot_phi}-\eqref{dot_r},
\begin{align}
\dfrac{\mathrm{d}r}{\mathrm{d}\varphi}=\pm r^2\sqrt{\dfrac{1}{b^2}-\dfrac{1}{r^2}\left(1-\dfrac{2M}{r}-\dfrac{4\pi\rho_cr_s^3}{r+r_s}\right)}.
\end{align}
By substituting $u\equiv1/r$, the equation above may be written as
\begin{align}
\label{trajectory}
\frac{\mathrm{d}u}{\mathrm{d}\varphi}=\sqrt{\frac{1}{b^2}-u^2\left[1-2Mu-\dfrac{4\pi\rho_cr_s^3u}{1+r_su}\right]}.
\end{align}
Via solving Eq.\eqref{trajectory} numerically using a ray tracing code, Fig.\ref{fig:trajectory_1} shows the trajectories of light ray. The black hole is shown in Fig.\ref{fig:trajectory_1} as a solid disk. Photons are caught by the black hole when the collision parameter $b<b_p$ (Region 3 in Fig.\ref{fig:Vph}). The blue lines in Fig.\ref{fig:trajectory_1} represent their paths. On the other hand, photons are redirected and moved towards infinity if $b>b_p$ (Region 1 in Fig.\ref{fig:Vph}), with their paths matching the yellow lines in Fig.\ref{fig:trajectory_1}. Photons circle the black hole indefinitely when $b=b_p$ (Region 2 in Fig.\ref{fig:Vph}). The red lines in Fig.\ref{fig:trajectory_1} represent their respective paths. It is clear that the red lines eventually form a circle, and the location of this circle exactly matches the black hole's photon sphere orbit. In Fig.\ref{fig:trajectory_1}, all incoming light rays are initialized to be parallel to the horizontal axis, which is a valid choice since the Schwarzschild-Hernquist spacetime is asymptotically flat. Moreover, it is impossible to ignore the substantial impact that the center density $\rho_c$ and the core radius $r_s$ of the Hernquist DM halo have on the photon paths.

\section{Observational features of thin disk emission}
\label{sec3}
In astrophysical settings, black holes are typically surrounded by large amounts of accreting matter. This makes the study of \replace{the shadows cast by}{the observational appearance of} Schwarzschild-Hernquist black holes particularly relevant from an observational perspective. \revision{We note that in the thin disk scenario discussed below, the boundary of the observed central brightness depression is determined by the emission profile and does not necessarily coincide with the critical curve at $b = b_p$, in contrast to the spherical accretion case. This distinction has been highlighted in recent works~\cite{gralla2019black,gralla2020lensing,macedo2024optical,paugnat2022photon}.}In this section we will investigate \replace{the shadows and rings}{the central brightness depression and ring structure} of a Schwarzschild-Hernquist black hole surrounded by a disk-like accretion disk that is optically and geometrically thin.

\subsection{Direct emission, lensing ring and photon ring}
According to Ref.~\cite{gralla2019black}, radiation with different impact parameters $b$ would cause the black hole to look differently to a distant observer when tracing rays backwards from an observer towards the black hole. We may use ray-tracing procedure to determine the optical appearance of a Schwarzschild-Hernquist black hole, by calculating the total deflection in azimuthal angle $\varphi$, for a photon traveling from the source to the observer. The number of orbits is then given by~\cite{gralla2019black,gralla2020lensing}
\begin{align}
\label{nb}
n(b)=\dfrac{\varphi}{2\pi},
\end{align}
Eq.\eqref{nb} depends on the impact parameter $b$. The number of orbits $n(b)$ is determined by how close the impact parameter b is to the critical value $b_p$. It also depends on the metric \eqref{metric} of the Schwarzschild-Hernquist black hole.

\begin{figure}[t]
\centering
\includegraphics[width=0.8\textwidth]{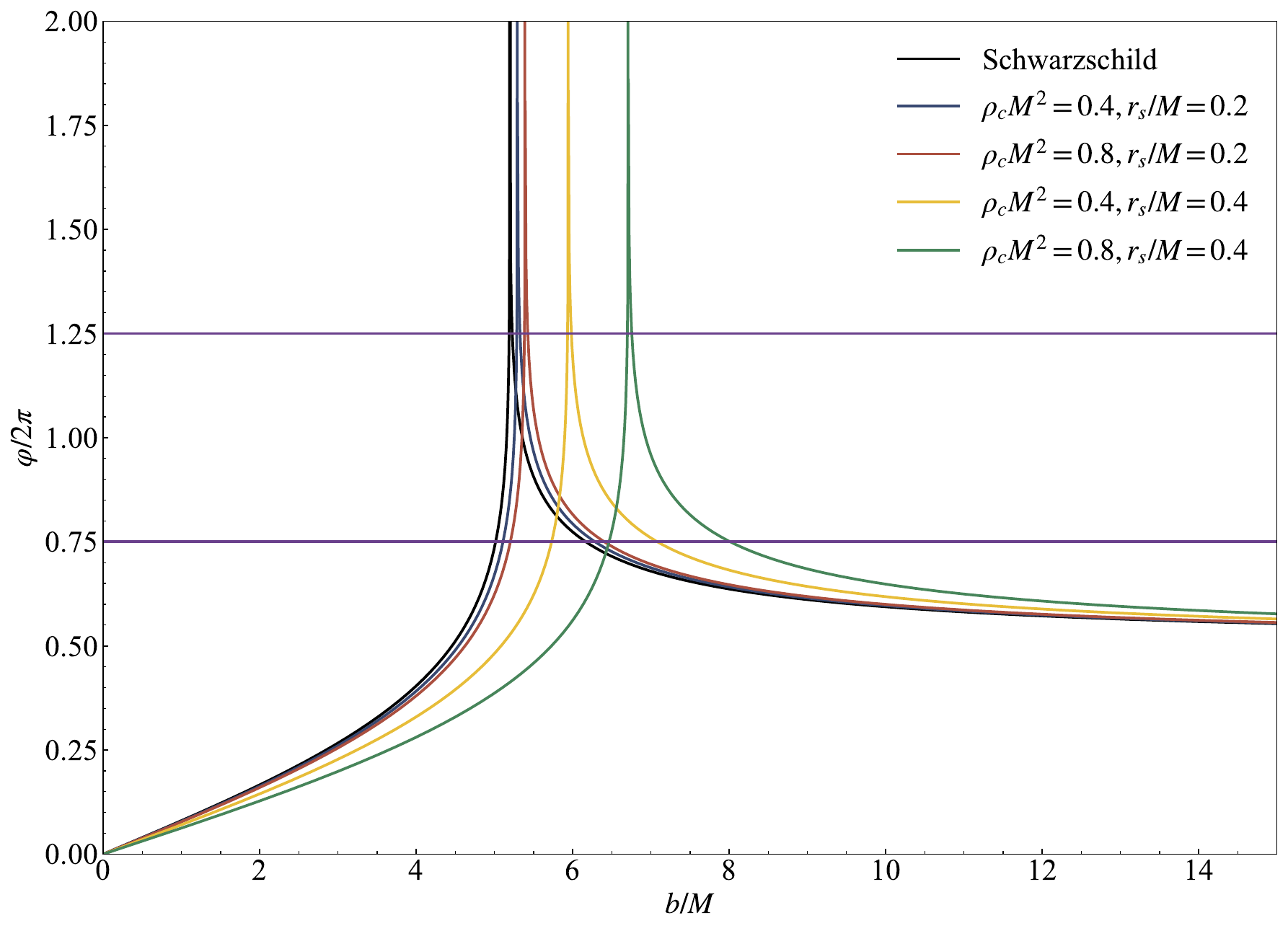}
\caption{The number of orbits $n$ as a function of the impact parameter $b$ is shown for various Hernquist DM parameters $\rho_c$ and $r_s$. Based on the number of plane intersections, the horizontal lines at $n=0.75$ and $n=1.25$ indicate three different observational zones: direct emission ($n<3/4$), the lensing ring ($3/4<n<5/4$) and the photon ring ($n>5/4$). The divergence of the curves corresponds to the critical impact parameter $b_p$.}
\label{fig:delphi}
\end{figure}

\begin{figure}[t]
\centering
\includegraphics[width=\textwidth]{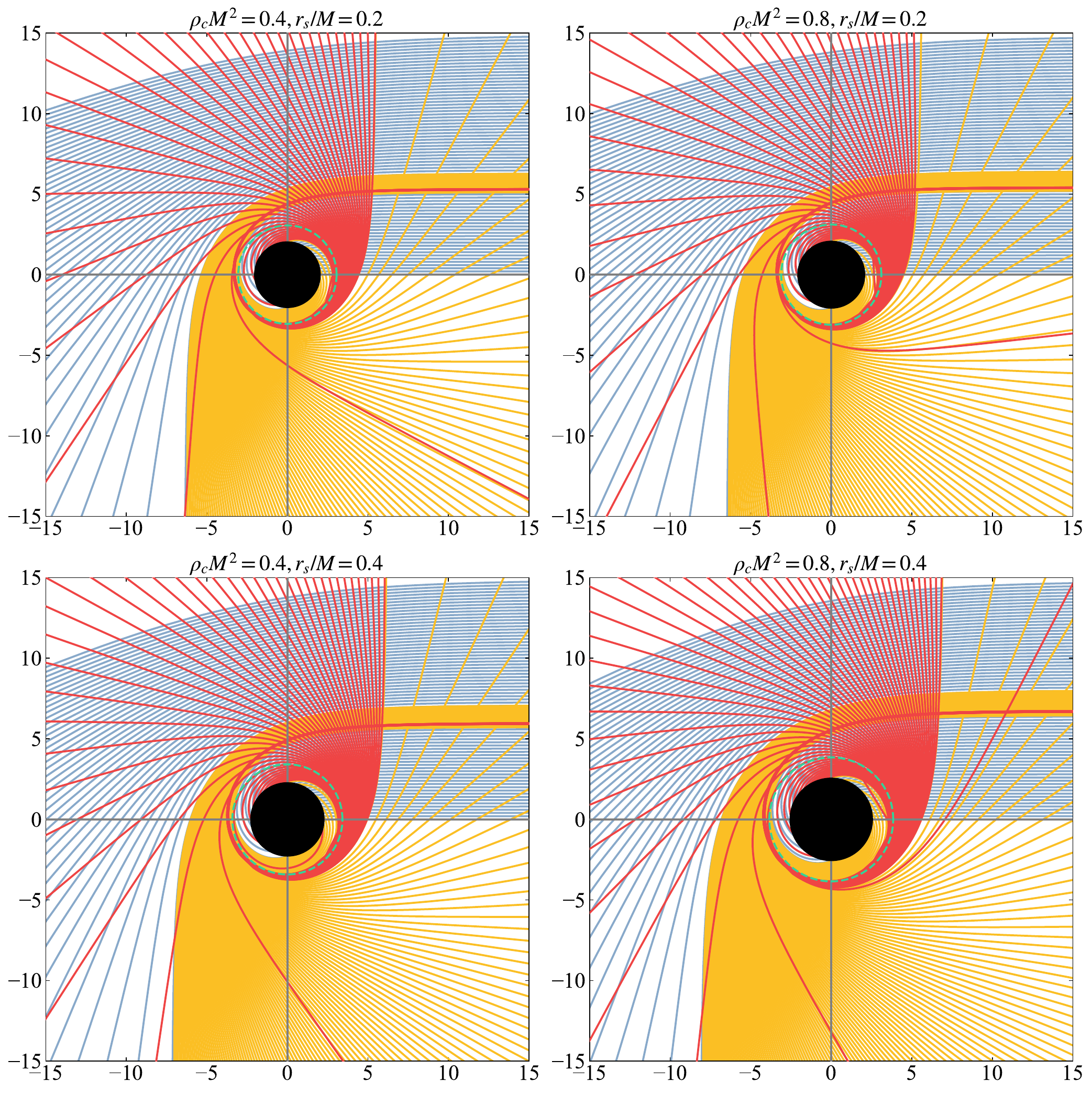}
\caption{Photon behaviour as a function of impact parameter $b$ in the Schwarzschild-Hernquist black hole. The blue lines represent trajectories contributing to direct emission ($n<3/4$), yellow lines constitute the lensing ring ($3/4<n<5/4$), and red lines mean the photon ring ($n>5/4$). The black disk and the green dashed circle indicates the event horizon and the photon sphere $r_p$, respectively.}
\label{fig:trajectory_2}
\end{figure}

\begin{table}[t]
\centering
\caption{The region of direct emission, lensing ring and photon ring is related to the impact parameters $b$ for different Hernquist DM parameters with $M=1$.}
\label{tab:orbit_ranges}
\tabcolsep=0.02\linewidth
\begin{tabular}{c|c|c|c}
\toprule[1.2pt]
& Type & $r_s/M=0.2$ & $r_s/M=0.4$ \\
\midrule[1.2pt]
\multirow{4}{*}{\makecell{$\rho_cM^2=0.4$}} 
& Direct emission & $b<5.10927$ or $b>6.28594$ & $b<5.73082$ or $b>7.08226$ \\
\cmidrule[0.8pt](lr){2-4}
& Lensing ring & \makecell{$5.10927<b<5.28566$\\or $5.32673<b<6.28594$} & \makecell{$5.73082<b<5.93441$\\or $5.98264<b<7.08226$} \\
\cmidrule[0.8pt](lr){2-4}
& Photon ring & $5.28566<b<5.32673$ & $5.93441<b<5.98264$ \\
\cmidrule[1.2pt](r){1-4}
\multirow{4}{*}{\makecell{$\rho_cM^2=0.8$}} 
& Direct emission & $b<5.20362$ or $b>6.40451$ & $b<6.46474$ or $b>8.01367$ \\
\cmidrule[0.8pt](lr){2-4}
& Lensing ring & \makecell{$5.20362<b<5.38374$\\or $5.42573<b<6.40451$} & \makecell{$6.46474<b<6.69895$\\or $6.75509<b<8.01367$} \\
\cmidrule[0.8pt](lr){2-4}
& Photon ring & $5.38374<b<5.42573$ & $6.69895<b<6.75509$ \\
\bottomrule[1.2pt]
\end{tabular}
\end{table}

In favour of the number of orbits $n(b)$, the emission can be categorized into three distinct zones~\cite{gralla2019black,gralla2020lensing,zeng2020influence,zeng2020shadows,zeng2022shadows,zeng2025holographic,li2021observational,li2021shadows}. These are lensing rings, photon rings, and direct emission rings corresponding to $n<3/4$, $3/4<n<5/4$, and $n>5/4$, respectively. These zones correspond to light rays that intersect the equatorial plane once, twice, or more than twice, respectively~\cite{gralla2019black,gralla2020lensing}. For various values of the parameters $\rho_c$ and $r_s$, Fig.\ref{fig:delphi} shows the functional dependence of $n(b)$ on the impact parameter $b$. The domains of direct emissions, lensing rings, and photon rings are shown by different colors. We have indicated the ring and the direct emission locations in Fig.\ref{fig:trajectory_2}. The function $n(b)$ diverges as bapproaches the critical value $b_p$, indicating the existence of unstable photon orbits, as shown in Fig.\ref{fig:delphi}. The curves for larger DM parameters $\rho_c$ and $r_s$ are shifted to the right, consistent with the increase in $b_p$ reported in Tab.\ref{tab:rh+rp+bp}. Tab.\ref{tab:orbit_ranges} lists the ranges of the impact parameter bcorresponding to direct emission, the lensing ring, and the photon ring for various values of $\rho_c$ and $r_s$. It is evident from the data that the range of the impact parameter $b$ for all emission types increases as the parameters increase. Furthermore, the total number of photon orbits $n(b)$ will peak inside a small area when the impact parameter $b$ gets closer to the critical condition $b\sim b_p$. The extremely narrow ranges for the photon ring listed in Tab.\ref{tab:orbit_ranges} correspond to the thin red band visible in Fig.\ref{fig:trajectory_2}. \replace{The angular size of the shadow also increases}{The angular radius of the critical curve (determined by $b_p$) also increases} with larger parameters. However, this enlargement does not directly translate to an increase in the brightness of the ring; the observed intensity depends on the complex interplay between the emission profile and the light bending.

\revision{We remark that an alternative classification scheme, widely adopted in recent studies~\cite{gralla2020lensing,johnson2020universal,
cardenas2023prediction,paugnat2022photon,vincent2022images,macedo2024optical}, labels photon trajectories by the number of half-turns $m$ performed around the black hole, rather than the number of orbits $n$ used in the present work. In this notation, $m = 0$ corresponds to direct emission, while $m = 1, 2, \ldots$ denote successive photon rings. The correspondence with our classification is straightforward: direct emission ($n < 3/4$) maps to $m = 0$, the lensing ring ($3/4 < n < 5/4$) to $m = 1$, and the photon ring ($n > 5/4$) to $m = 2$. We retain the notation of Ref.~\cite{gralla2019black} throughout this paper for internal consistency.}


\subsection{Transfer functions}
We now study the emission from a thin disk viewed in a face-on orientation. In this case, we suppose that the thin disk is located on the Schwarzschild-Hernquist black hole's equatorial plane and that its emission is isotropic in the rest frame of the emitting material. The specific intensity measured by a stationary observer at infinity, neglecting absorption and scattering, is given by~\cite{zeng2020influence,zeng2020shadows,zeng2022shadows,zeng2025holographic,li2021observational,li2021shadows}
\begin{align}
I_{\text{obs}}(r)=\left(1-\dfrac{2M}{r}-\dfrac{4\pi\rho_cr_s^3}{r+r_s}\right)^{3/2}I_{\text{e}}(r).
\end{align}
The total observed intensity, which integrated over all frequencies, is obtained by integrating~\cite{meng2023images}
\begin{align}
\label{I_o}
I\left(r\right)
&=\int I_{\text{obs}}(r)\mathrm{d}\nu\notag\\
&=\left(1-\dfrac{2M}{r}-\dfrac{4\pi\rho_cr_s^3}{r+r_s}\right)^2I_{\text{emi}}(r),
\end{align}
where the total emission intensity is given by
\begin{align}
I_{\text{emi}}(r)=\int I_{\text{e}}(r)\mathrm{d}\nu_{\text{e}}.
\end{align}
Since a light ray may intersect the accretion disk multiple times, each intersection contributes to the observed intensity. Therefore, the total observed intensity is a sum over these intersections:
\begin{align}
I\left(r\right)=\left(1-\dfrac{2M}{r}-\dfrac{4\pi\rho_cr_s^3}{r+r_s}\right)^{2}\sum_{n}I_{\text{emi}}(r)|_{r=r_{n}(b)}.
\end{align}

\begin{figure}[t]
\centering
\includegraphics[width=\textwidth]{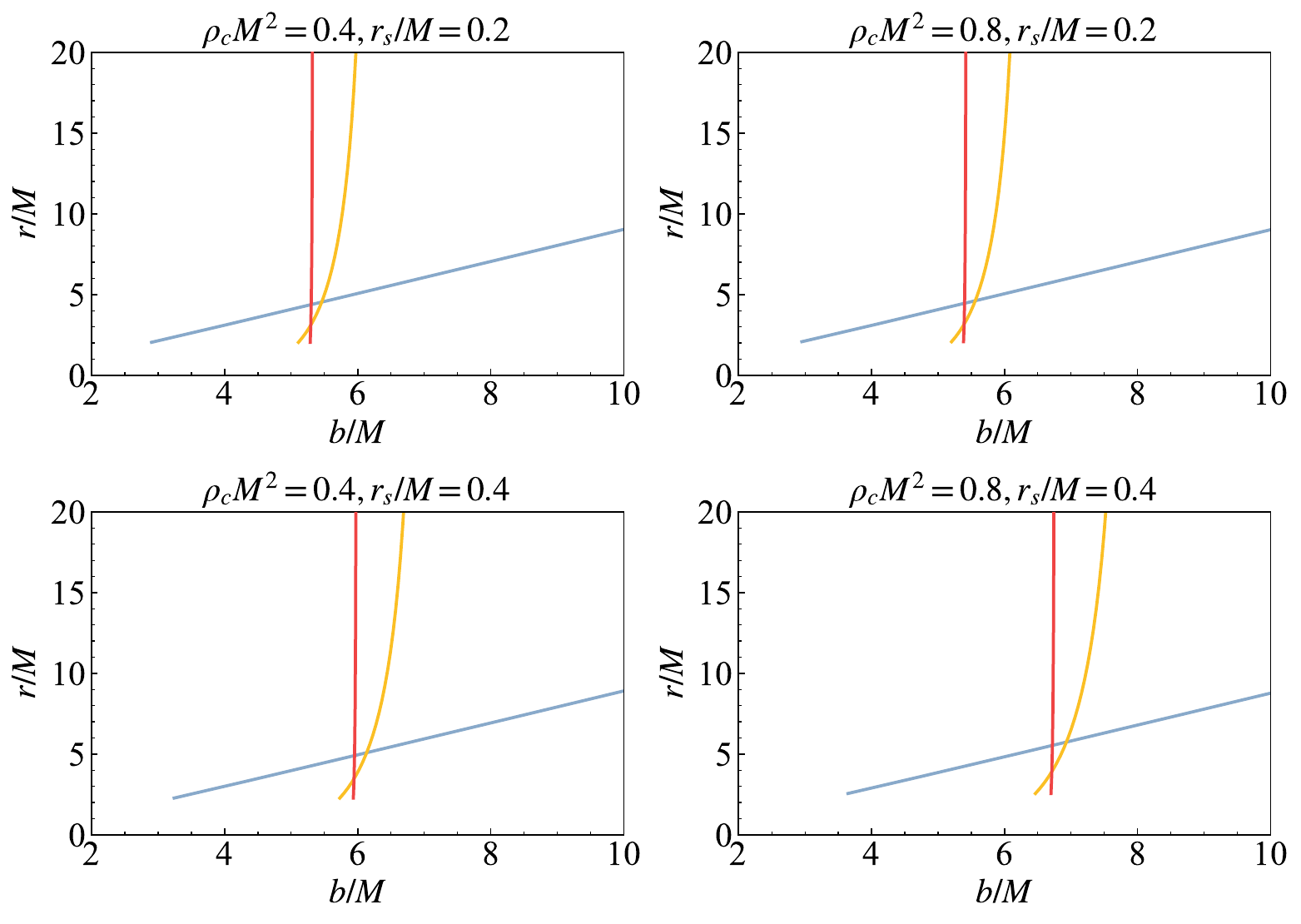}
\caption{The first three transfer functions for the Schwarzschild-Hernquist black hole. The curves correspond to the first (blue), second (yellow) and third (red) intersections with the emission,
respectively. We set $M=1$ here.}
\label{fig:transfer}
\end{figure}

The transfer function $r_n(b)$ encodes the radial coordinate on the disk of the $n$-th intersection of a light ray with impact parameter $b$. Furthermore, the transfer function's slope, often called the demagnification factor, is $\mathrm{d}r/\mathrm{d}b$, and it determines the magnification of the image of the disk~\cite{gralla2019black,li2021observational,fathi2023observational,he2022shadow,shi2024shadow,zeng2020influence}. In Fig.\ref{fig:transfer}, the transfer function for direct emission $(n=1)$ is shown by the blue line. Since the direct imaging profile is the redshifted source profile, the slope $\mathrm{d}r/\mathrm{d}b$ of the blue line remains approximately constant and close to 1 despite changes in the Hernquist DM parameters. For the lensing ring $(n=2)$, the transfer function is shown by the yellow line. The slope of the second transfer function rises to a very high value when the impact parameter b gets closer to the critical condition $b\sim b_p$. For the photon ring $(n=3)$, the red line represents the transfer function. The slope of the red line approaches infinity, indicating extreme demagnification of the disk's image. This implies that direct emission dominates the total observed flux, with the photon ring and lensing ring contributing only a small fraction. Notably, the contribution of transfer functions with $n\geq4$ has been discovered to provide some interferometric features~\cite{johnson2020universal}, but it is insignificant and negligible. Therefore, we consider only the first three transfer functions $n=1,2,3$.


\subsection{Observational features}
To compute the observational appearance of the thin disk, we employ three phenomenological emissivity models~\cite{wang2022optical,yang2023shadow,li2021shadows,zeng2020influence,zeng2020shadows,zeng2022shadows,zeng2025holographic}. \revision{These three emission profiles follow the toy-model introduced by Gralla~\cite{gralla2019black}, which has been widely adopted in the black hole shadow literature.} It is important to note that these models are simplified representations and are strategically chosen to represent a diverse range of plausible astrophysical scenarios from efficient, geometrically thin disks to inefficient, thick flows~\cite{meng2023images,gan2021photon,guo2022gravitational,chen2022appearance}. This approach allows us to systematically explore how the observational signatures of the Hernquist DM halo depend on the spatial distribution of the emitting matter. While the precise quantitative features of the image, such as peak brightness, are model-dependent, we expect the qualitative findings, particularly the scaling of \replace{the shadow}{the critical curve} and photon ring diameters with DM parameters, to be robust across different emission profiles.

\textbf{Model 1} is designed to represent a standard, geometrically thin, optically thick accretion disk, \replace{analogous to}{inspired by} the classical Novikov-Thorne model~\cite{shakura1973black,novikov1973astrophysics,page1974disk}. The emissivity is sharply peaked at the innermost stable circular orbit (ISCO), which is the radius where gravitational energy is most efficiently released, and follows a rapid power-law decay outwards. This profile is characteristic of radiatively efficient accretion scenarios, often seen in luminous active galactic nuclei (AGN)~\cite{yuan2014hot,koratkar1999ultraviolet}.
\begin{align}
\label{Iemi_2nd}
I_{\text{emi}}(r)=
\left\{
\begin{array}{ll}
\left[\dfrac{1}{r-\left(r_{\text{ISCO}}-1\right)}\right]^{2},&r>r_{\text{ISCO}}\\
0,&r\leq r_{\text{ISCO}}
\end{array}
\right..
\end{align}

\textbf{Model 2} serves as an exploratory case to probe the observational appearance when the emission peak is located extremely close to the unstable photon orbits at $r_p$. While not a standard astrophysical scenario, this profile could hypothetically arise from localized energy release, such as magnetic reconnection events occurring \replace{near the photon sphere}{in the innermost black hole magnetosphere}~\cite{ripperda2022black,sironi2020kinetic}. Its primary purpose is to test the most extreme gravitational lensing effects on an emission source situated at the boundary of photon capture.
\begin{align}
I_{\text{emi}}(r)=
\left\{
\begin{array}{ll}
\left[\dfrac{1}{r-\left(r_{p}-1\right)}\right]^{3},&r>r_{p}\\
0,&r\leq r_{p}
\end{array}
\right..
\end{align}

\textbf{Model 3} is chosen to represent a physically distinct scenario: a geometrically thick, optically thin, and radiatively inefficient accretion flow, such as an Advection-Dominated Accretion Flow (ADAF)~\cite{narayan1994advection,narayan1995advection,yuan2014hot}. This type of flow is believed to be present in low-luminosity AGN like $\text{M87}^*$ and $\text{SgrA}^*$~\cite{yuan2003nonthermal,EHT1}. The emissivity profile is consequently broader, with a more gradual decay, reflecting emission that is spread over a much wider radial range rather than being concentrated at the inner edge.
\begin{align}
\label{Iemi_mod}
I_{\text{emi}}(r)=
\left\{
\begin{array}{ll}
\dfrac{\frac{\pi}{2}-\tan^{-1}\left[r-\left(r_{\text{ISCO}}-1\right)\right]}{\frac{\pi}{2}+\tan^{-1}(r_{p})},&r>r_h\\
0,&r\leq r_h
\end{array}
\right..
\end{align}

\begin{figure}[t]
\centering
\begin{subfigure}[t]{0.33\textwidth}
\centering
\includegraphics[width=\textwidth]{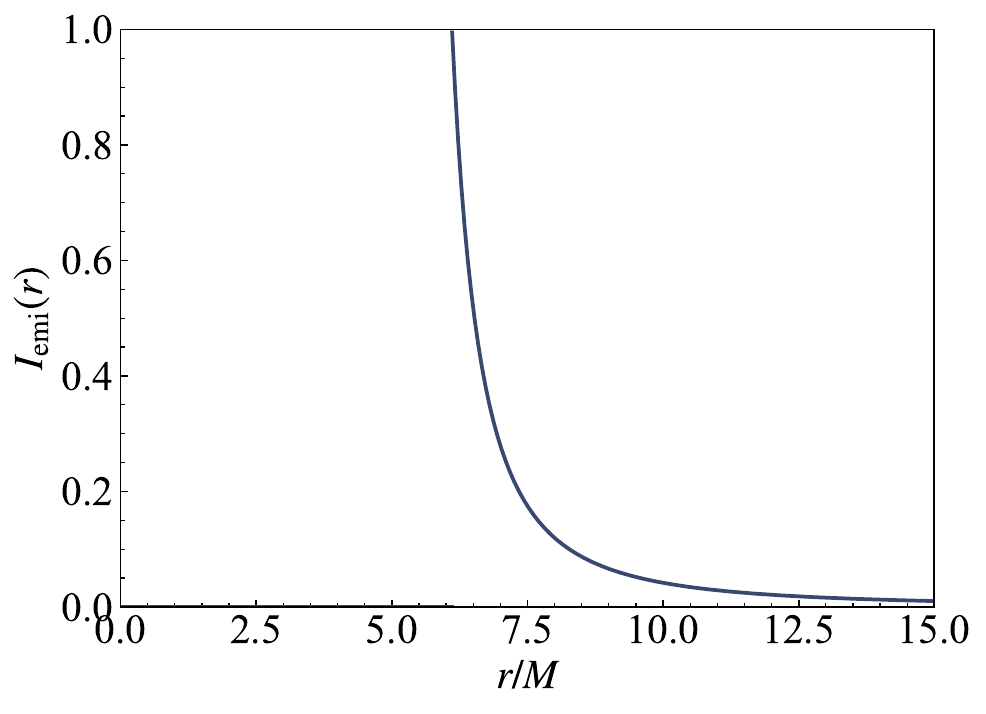}
\end{subfigure}
\begin{subfigure}[t]{0.33\textwidth}
\centering
\includegraphics[width=\textwidth]{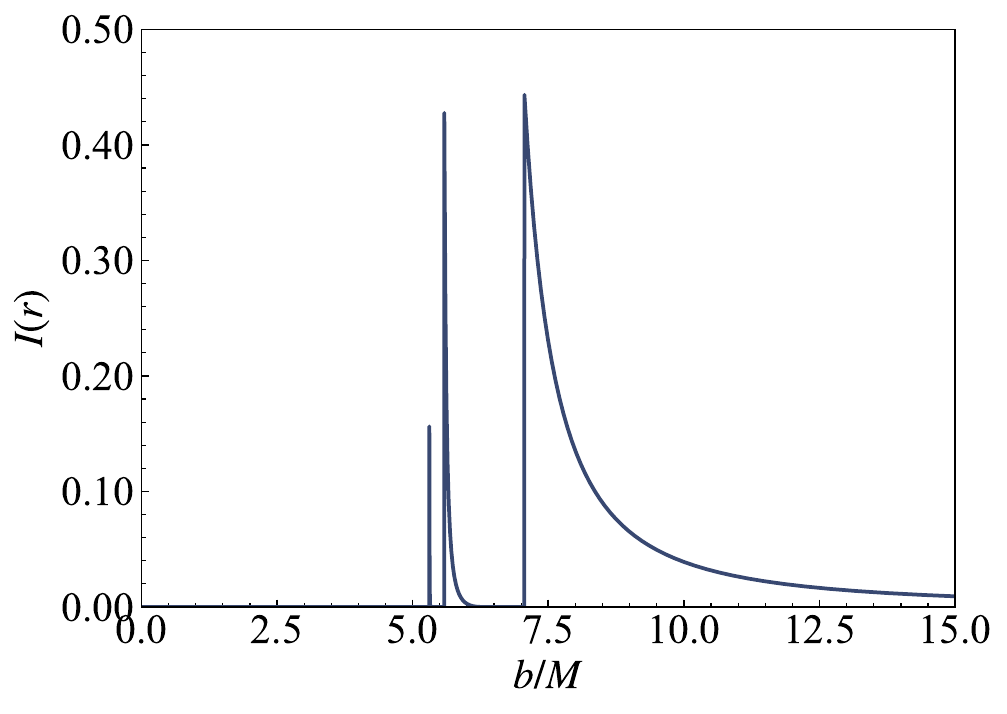}
\end{subfigure}
\begin{subfigure}[t]{0.3\textwidth}
\centering
\includegraphics[width=\textwidth]{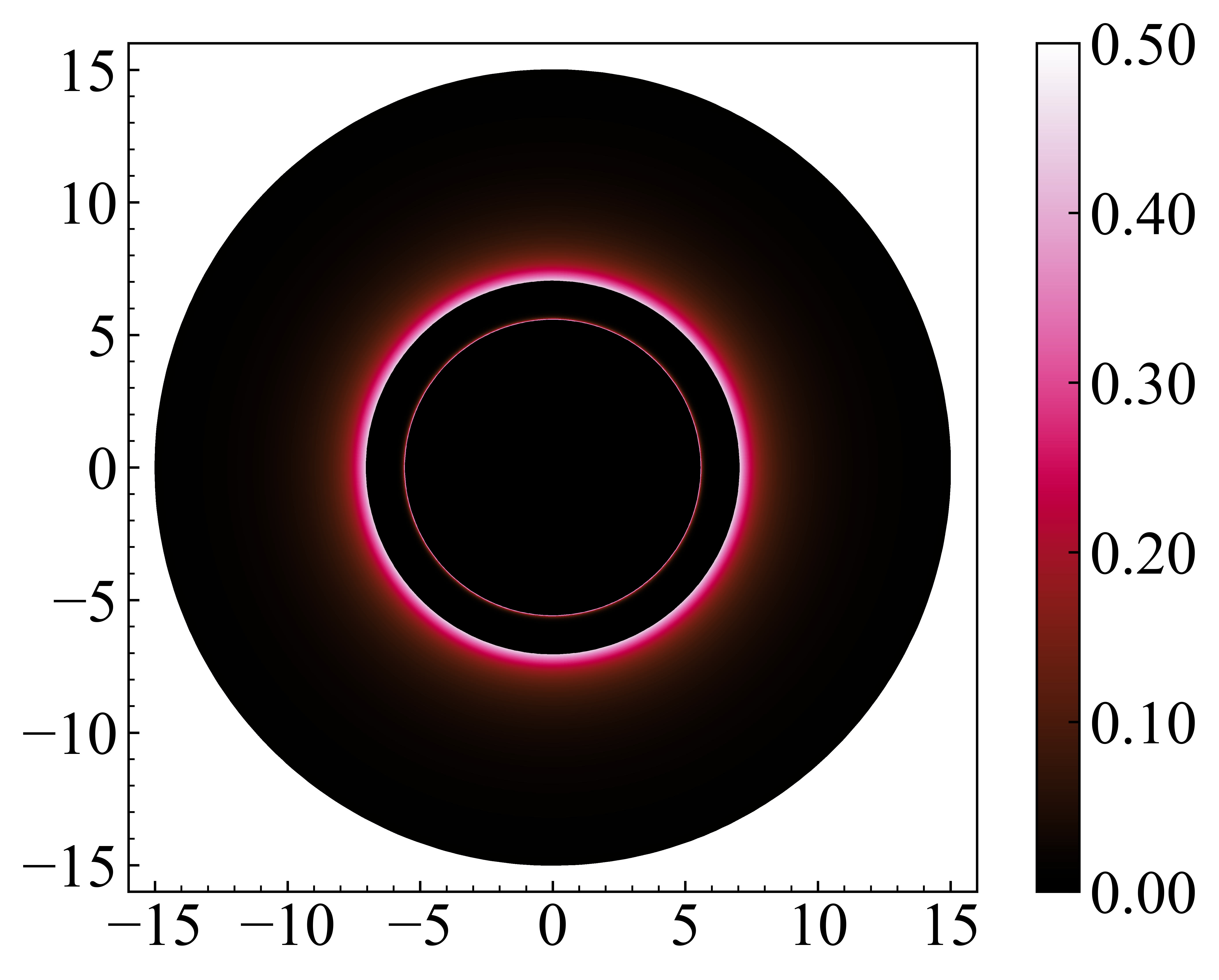}
\end{subfigure}
\begin{subfigure}[t]{0.33\textwidth}
\centering
\includegraphics[width=\textwidth]{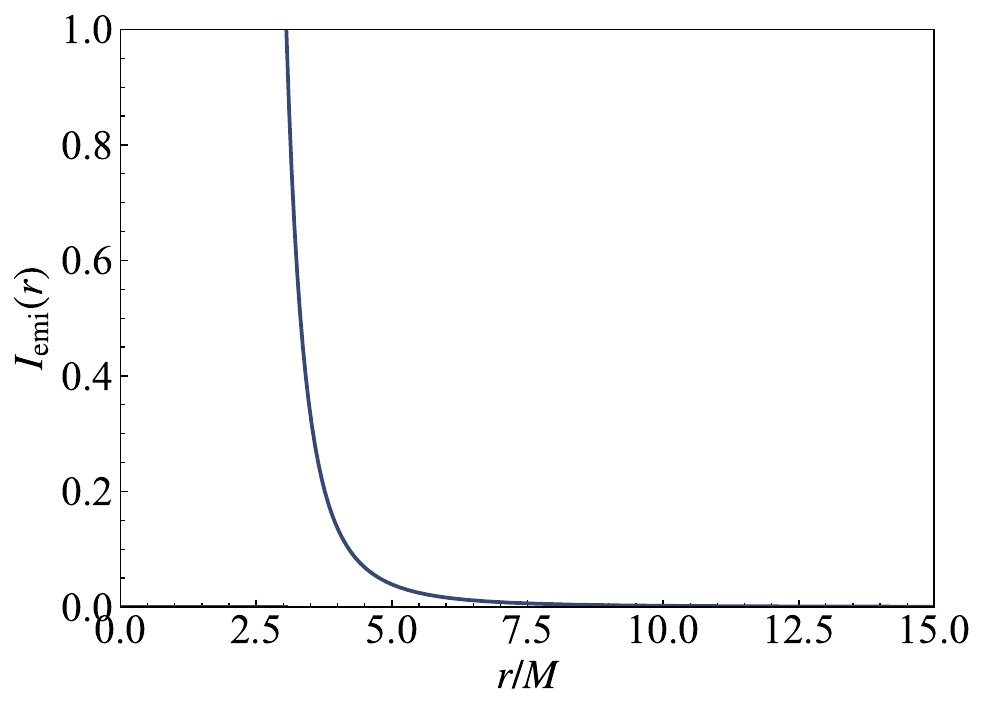}
\end{subfigure}
\begin{subfigure}[t]{0.33\textwidth}
\centering
\includegraphics[width=\textwidth]{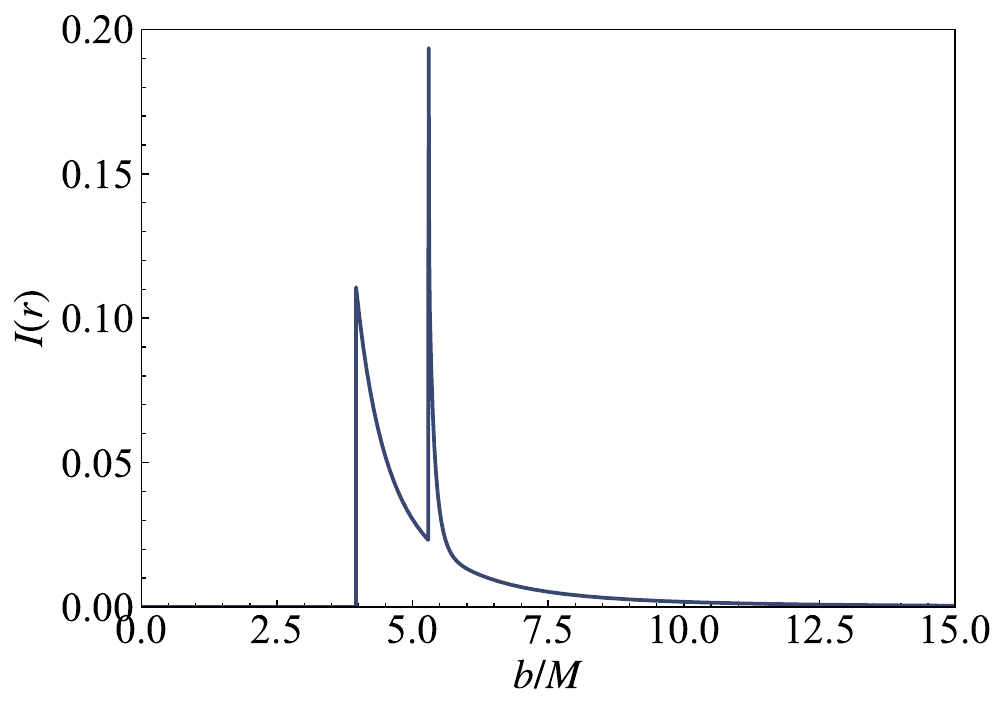}
\end{subfigure}
\begin{subfigure}[t]{0.3\textwidth}
\centering
\includegraphics[width=\textwidth]{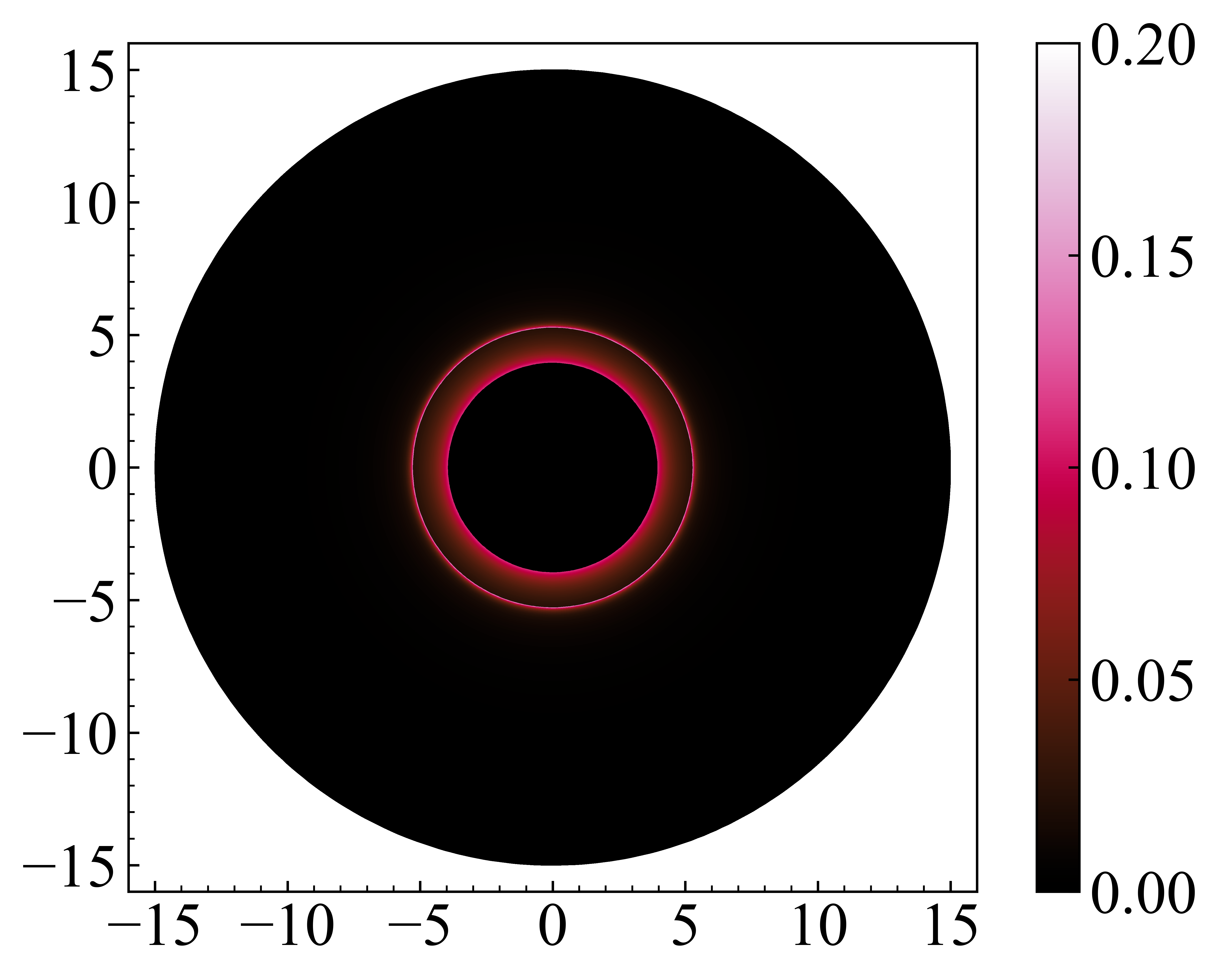}
\end{subfigure}
\begin{subfigure}[t]{0.33\textwidth}
\centering
\includegraphics[width=\textwidth]{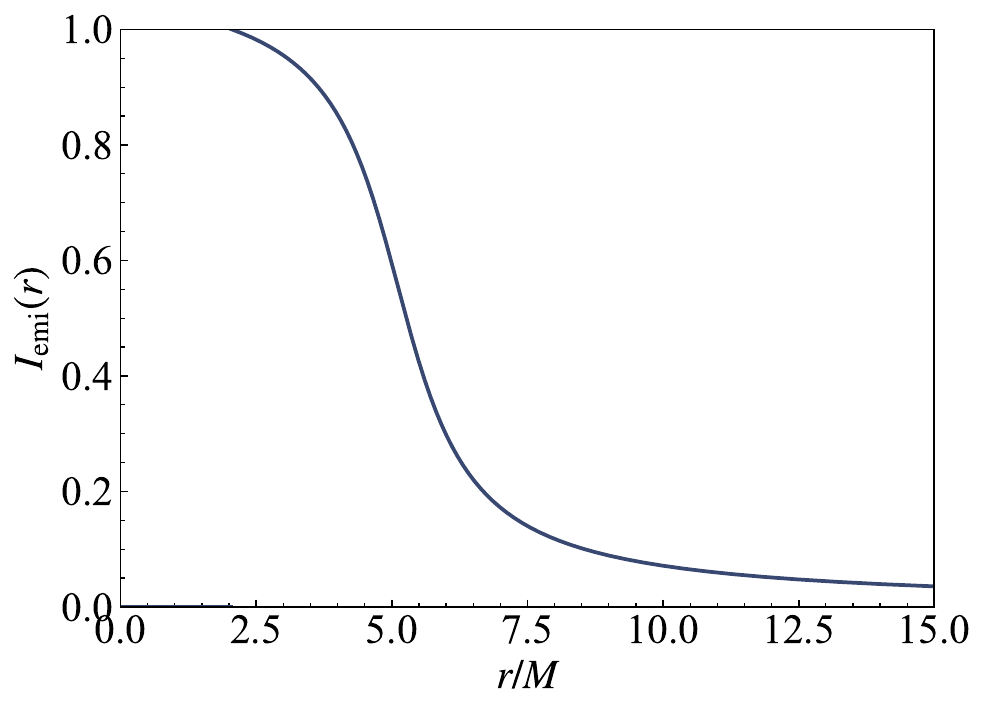}
\end{subfigure}
\begin{subfigure}[t]{0.33\textwidth}
\centering
\includegraphics[width=\textwidth]{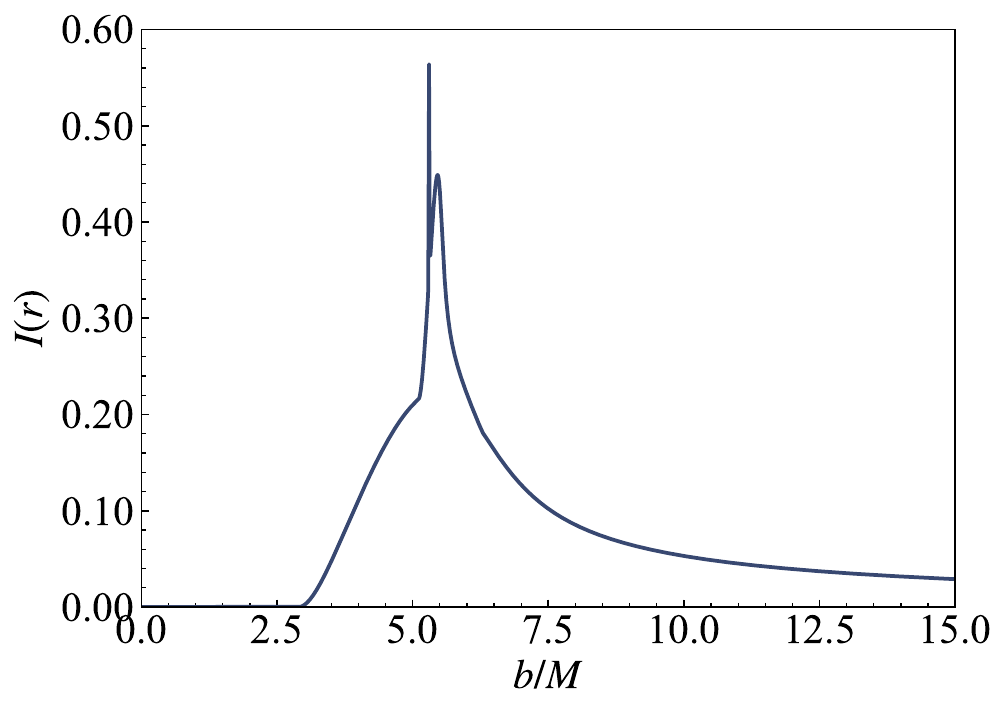}
\end{subfigure}
\begin{subfigure}[t]{0.3\textwidth}
\centering
\includegraphics[width=\textwidth]{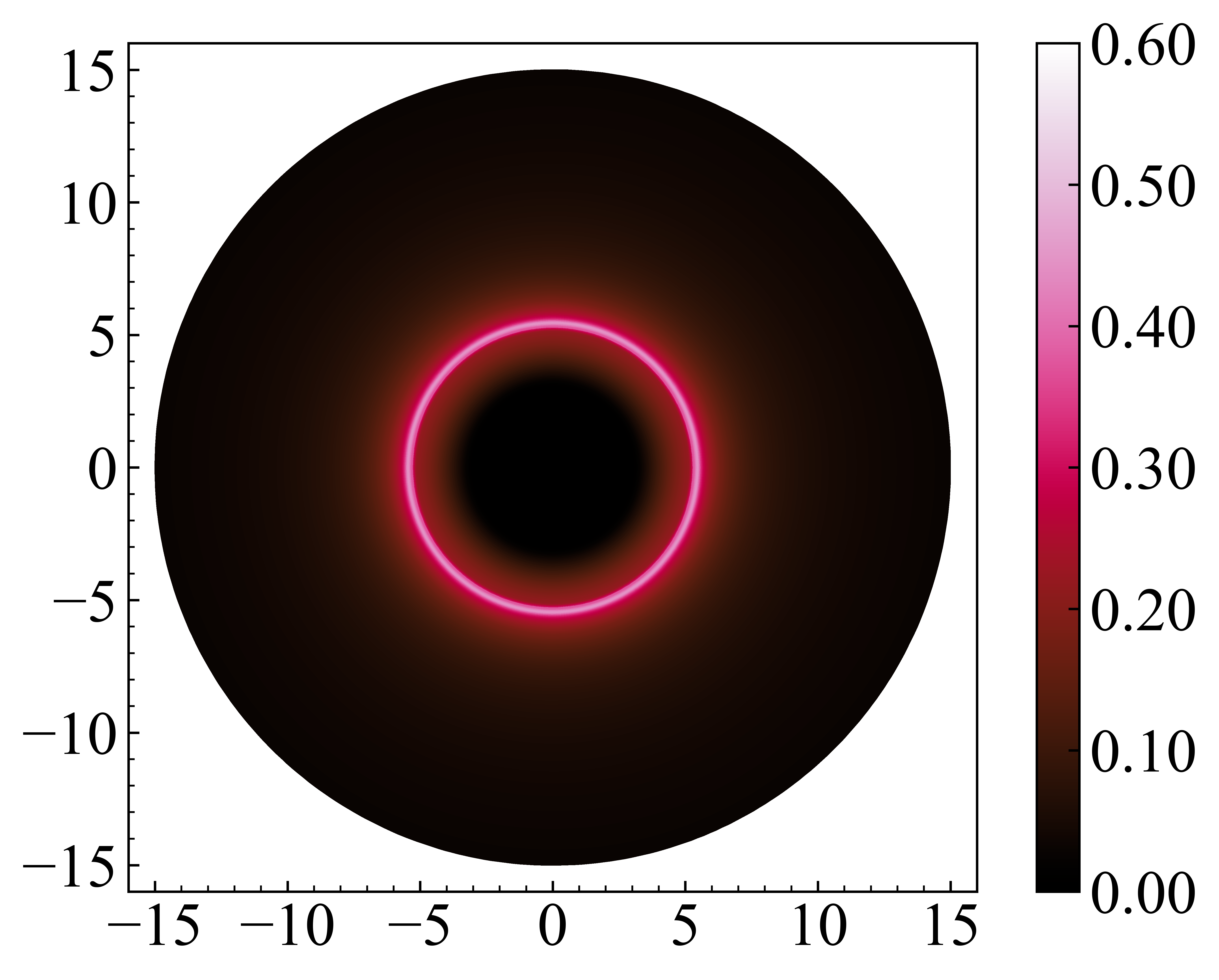}
\end{subfigure}
\caption{Observational signatures of the thin disk surrounding the Schwarzschild-Hernquist black hole with various emission profiles in case of $\rho_cM^2=0.4$ and $r_s/M=0.2$ for three different emissivity models. The rows correspond to \textbf{Model 1} (Top), \textbf{Model 2} (Middle), and \textbf{Model 3} (Bottom). Left: The emissivity profile $I_{\text{emi}}(r)$ as a function of radius $r/M$. Center: The observed specific intensity $I_{\text{obs}}(b)$ as a function of the impact parameter $b/M$. Right: The two-dimensional optical appearance with color bars, which mean the intensities.}
\label{fig:case1}
\end{figure}

\begin{figure}[t]
\centering
\begin{subfigure}[t]{0.33\textwidth}
\centering
\includegraphics[width=\textwidth]{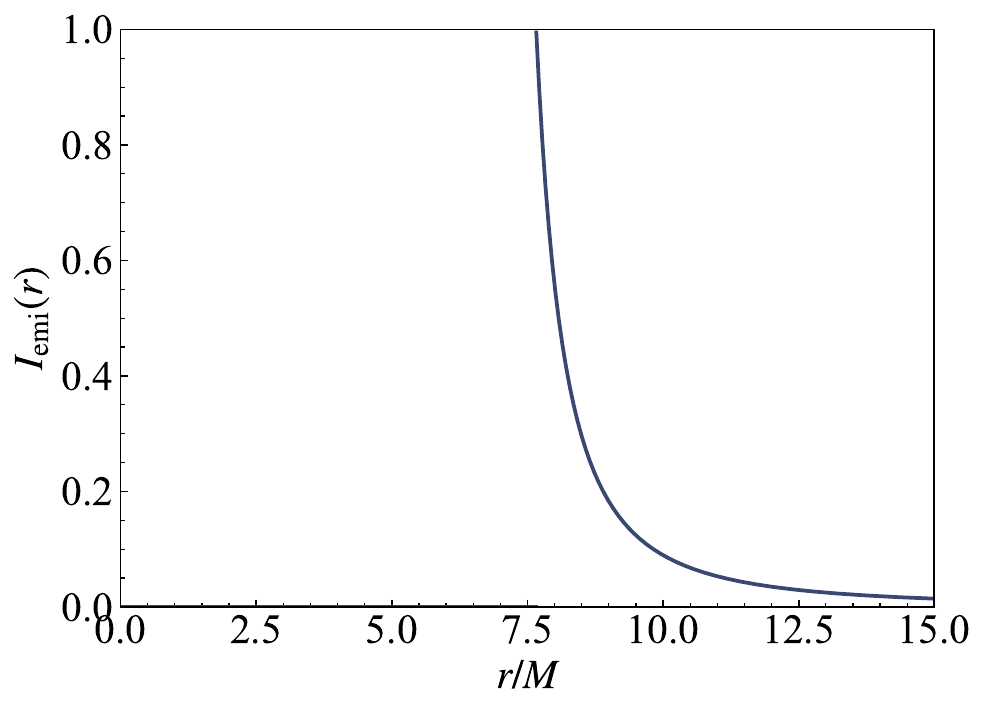}
\end{subfigure}
\begin{subfigure}[t]{0.33\textwidth}
\centering
\includegraphics[width=\textwidth]{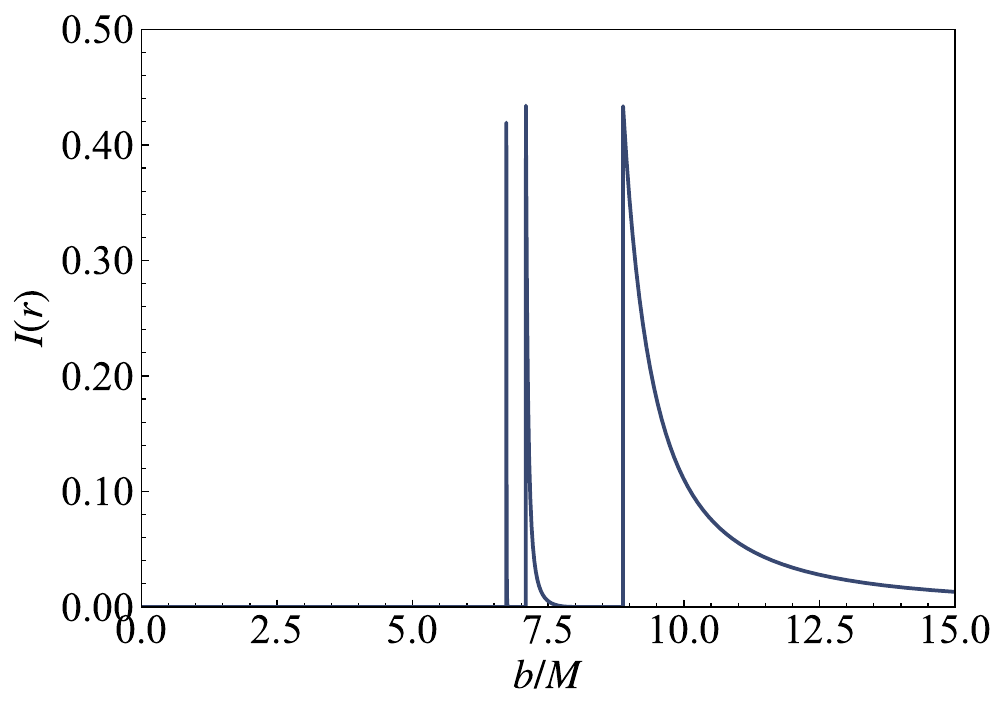}
\end{subfigure}
\begin{subfigure}[t]{0.3\textwidth}
\centering
\includegraphics[width=\textwidth]{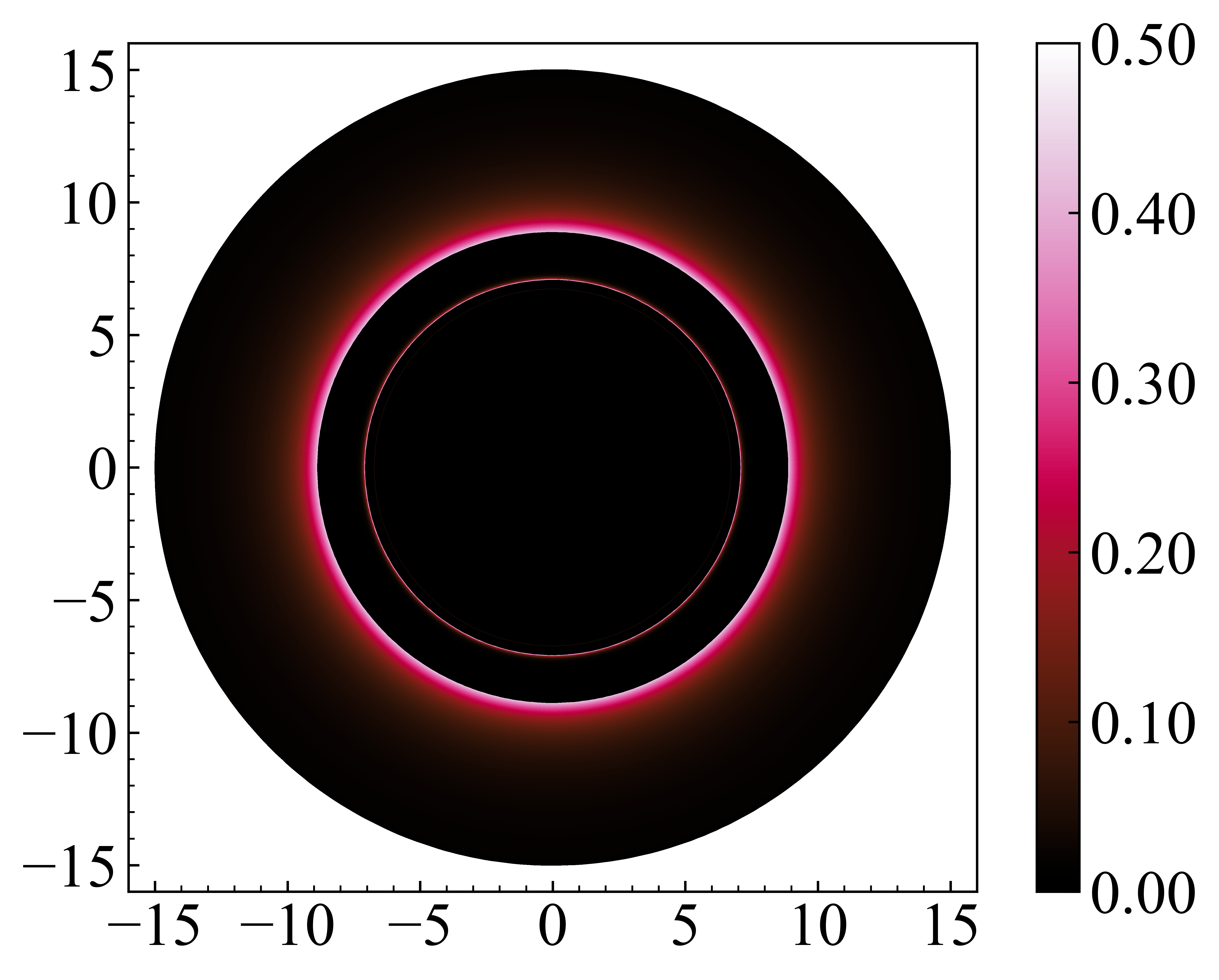}
\end{subfigure}
\begin{subfigure}[t]{0.33\textwidth}
\centering
\includegraphics[width=\textwidth]{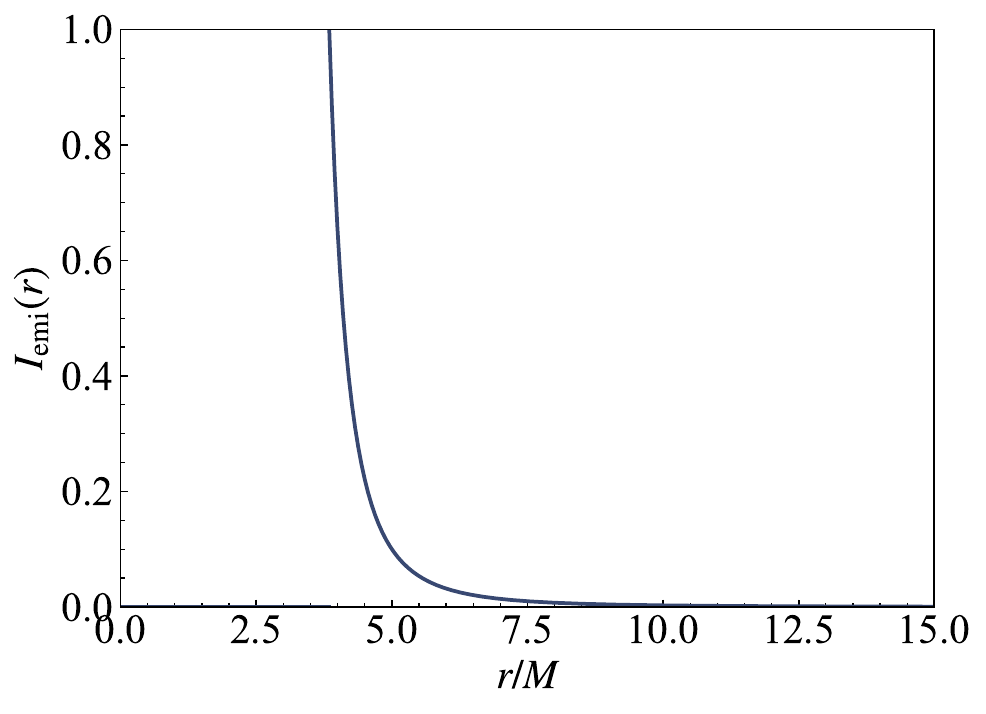}
\end{subfigure}
\begin{subfigure}[t]{0.33\textwidth}
\centering
\includegraphics[width=\textwidth]{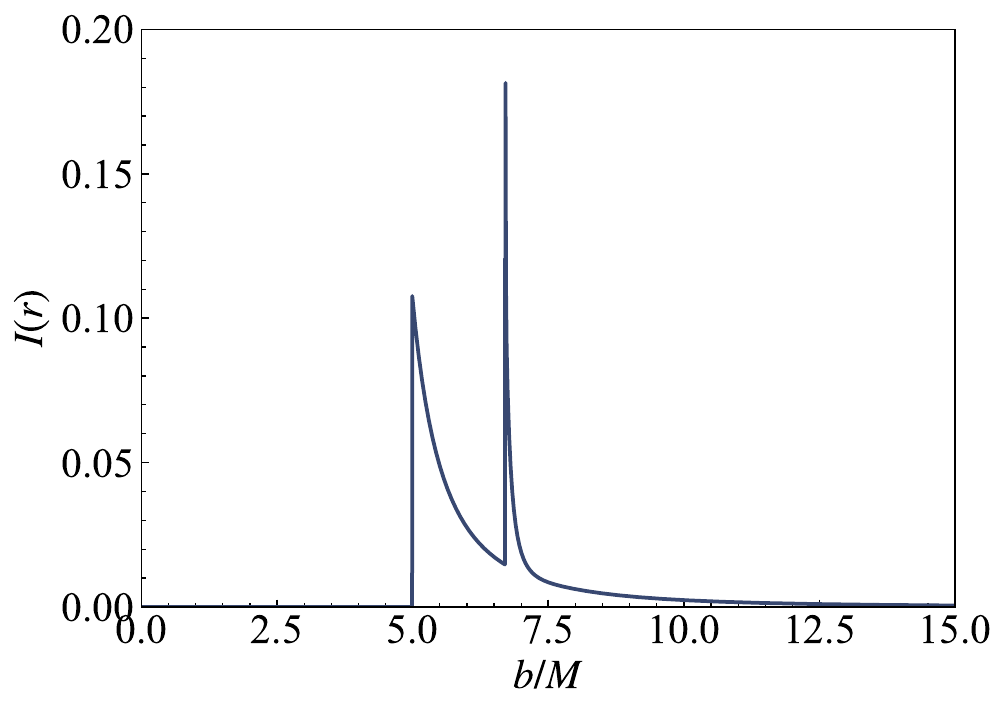}
\end{subfigure}
\begin{subfigure}[t]{0.3\textwidth}
\centering
\includegraphics[width=\textwidth]{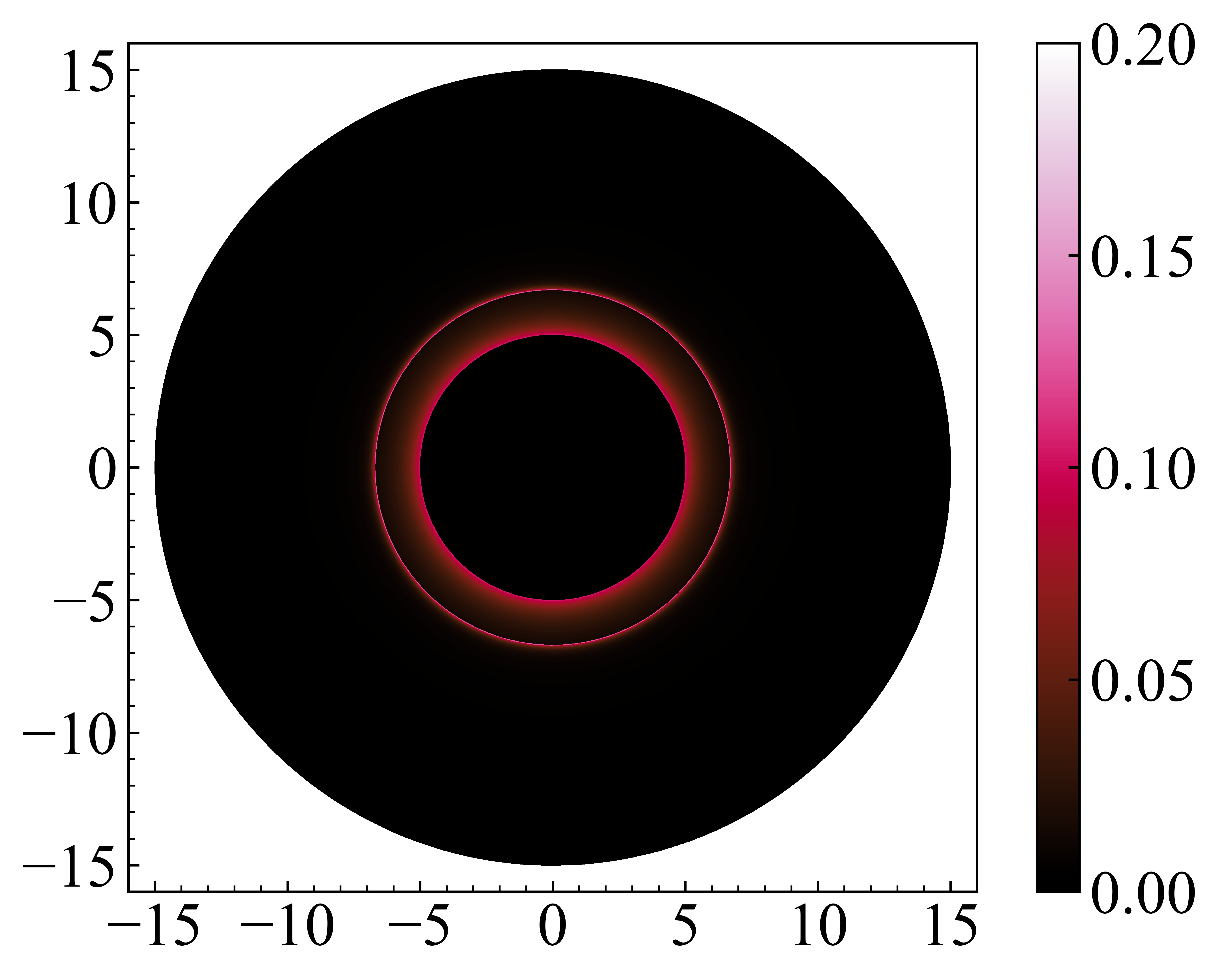}
\end{subfigure}
\begin{subfigure}[t]{0.33\textwidth}
\centering
\includegraphics[width=\textwidth]{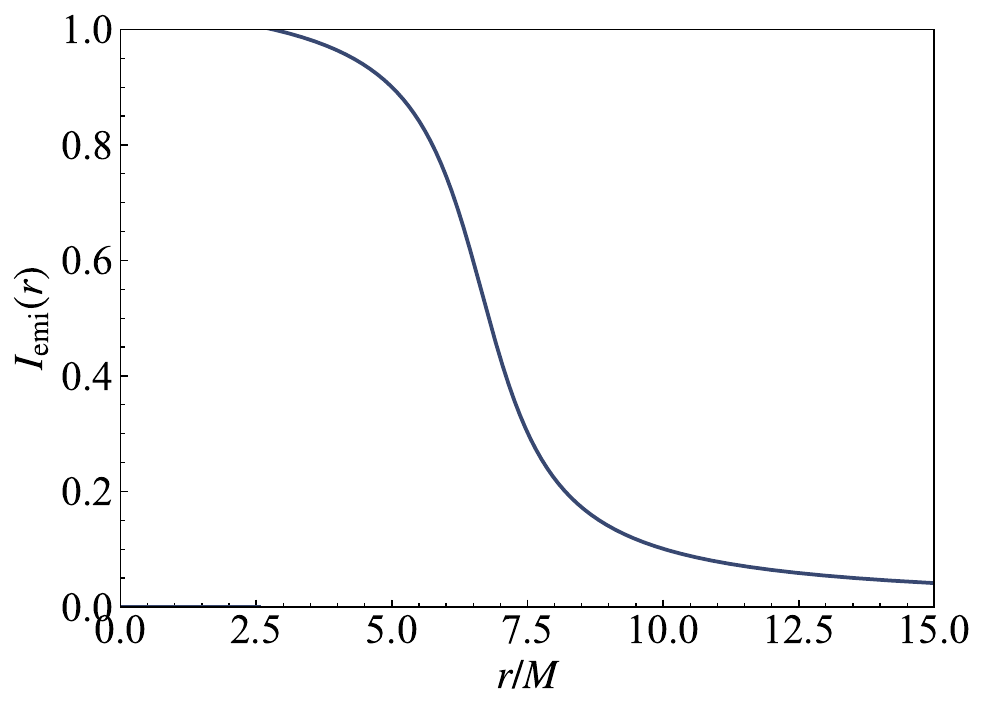}
\end{subfigure}
\begin{subfigure}[t]{0.33\textwidth}
\centering
\includegraphics[width=\textwidth]{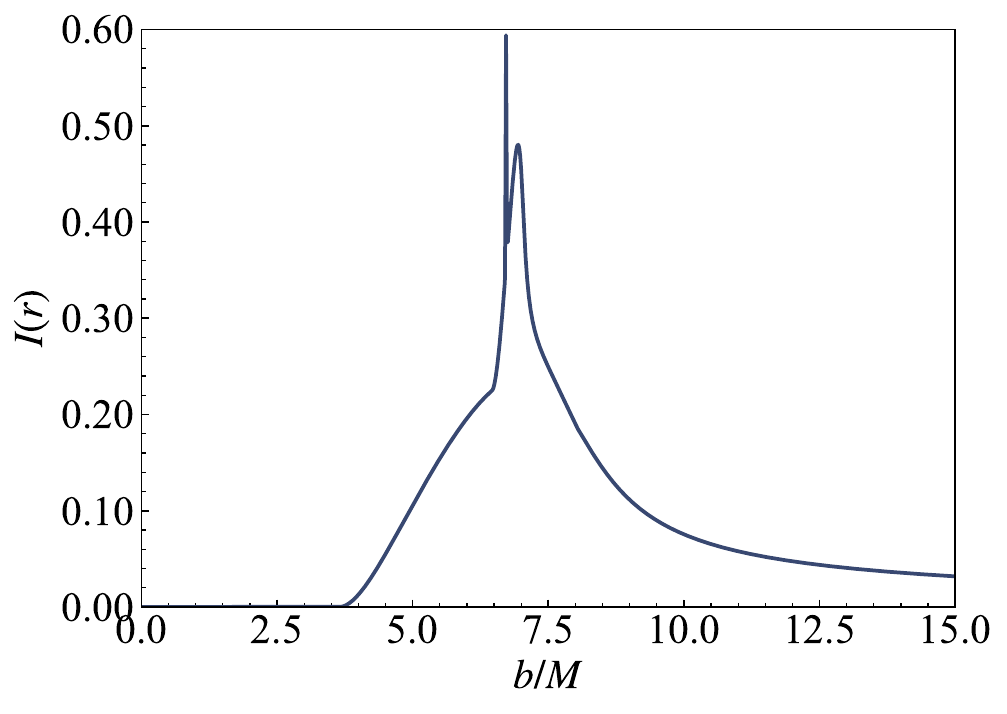}
\end{subfigure}
\begin{subfigure}[t]{0.3\textwidth}
\centering
\includegraphics[width=\textwidth]{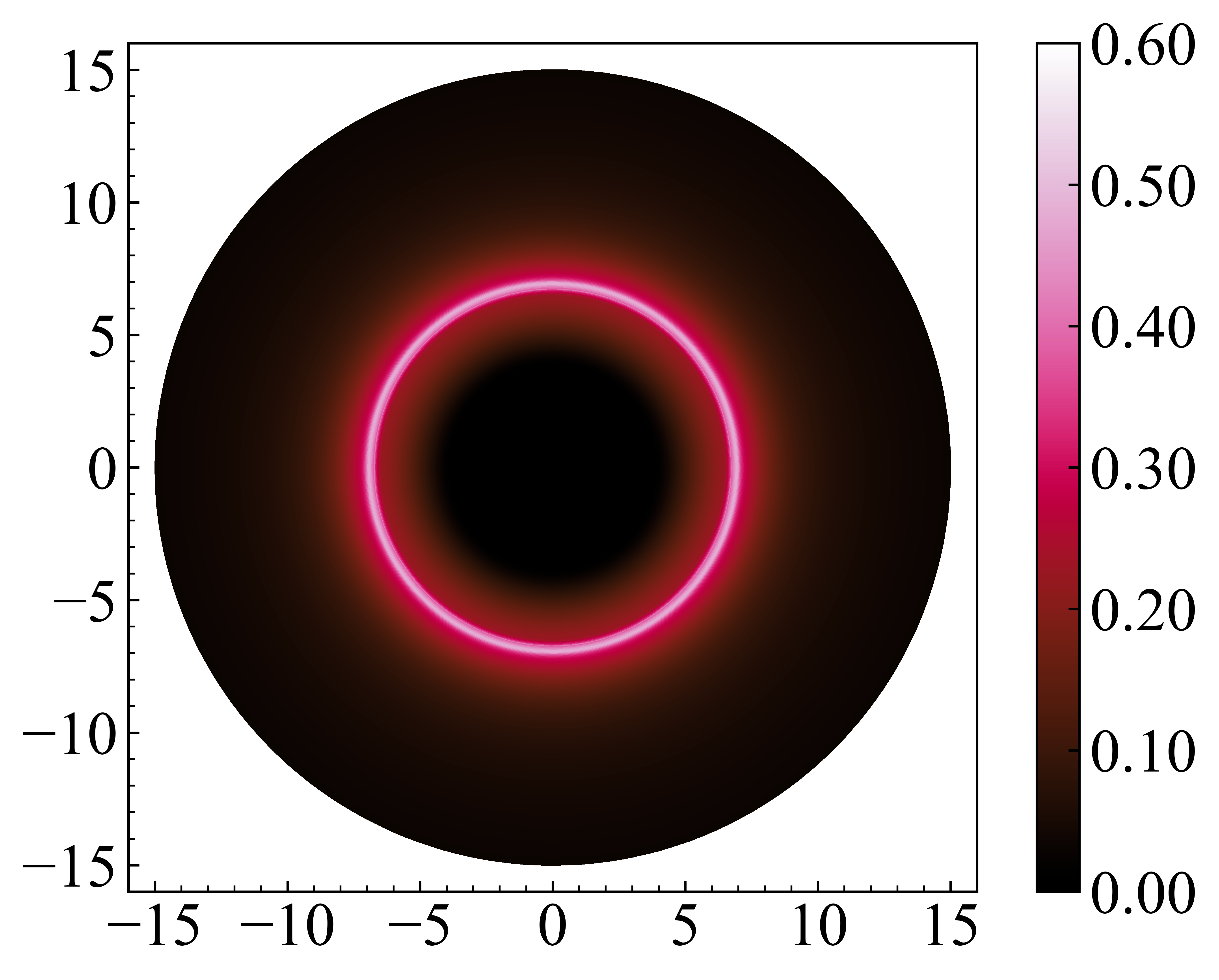}
\end{subfigure}
\caption{Observational signatures of the thin disk surrounding the Schwarzschild-Hernquist black hole with various emission profiles in case of $\rho_cM^2=0.8$ and $r_s/M=0.4$ for three different emissivity models. The rows correspond to \textbf{Model 1} (Top), \textbf{Model 2} (Middle), and \textbf{Model 3} (Bottom). Left: The emissivity profile $I_{\text{emi}}(r)$ as a function of radius $r/M$. Center: The observed specific intensity $I_{\text{obs}}(b)$ as a function of the impact parameter $b/M$. Right: The two-dimensional optical appearance with color bars, which mean the intensities.}
\label{fig:case2}
\end{figure}

We demonstrate the results for two representative parameter sets: $\rho_cM^2=0.4,r_s/M=0.2$ (Fig.\ref{fig:case1}) and $\rho_cM^2=0.8,r_s/M=0.4$ (Fig.\ref{fig:case2}). The emission profiles of the three models are shown in the left column of Figs.\ref{fig:case1} and \ref{fig:case2} from top to bottom as \textbf{Model 1}, \textbf{Model 2} and \textbf{Model 3}. The observed specific intensity, which is dependent on the impact parameter $b$, is given in the centre column. Additionally, the right column provides the two-dimensional observed image.

The first row of the left column in Figs.\ref{fig:case1} and \ref{fig:case2} shows that the emissivity peaks near $r_{\text{ISCO}}$ and decays rapidly to zero with increasing $r$. In this instance, the photon sphere lies inside the emission zone. At the same time, two distinct peaks in the observed intensity profile, corresponding to the lensing ring and photon ring, can be identified, each of which can be separately differentiated due to the gravitational lensing effect during observation. However, compared to the direct emission peak, the photon sphere and the lensing ring occupy geometrically smaller observational areas and, depending on the emission profile, exhibit lower observational intensity peaks. Consequently, for this specific accretion scenario, the total observed flux is dominated by direct emission, whereas the photon sphere contributes negligibly to the total flux and lensing ring emission contributes only a small fraction. The right-hand column displays the primary optical appearance, which is caused by direct emission, and the photon ring is barely visible.

The emission peak in the left column of the second row in Figs.\ref{fig:case1} and \ref{fig:case2} is shifted outward. According to the results in the centre column, the measured intensity will rise to the first peak due to direct emission and then show a trend of progressive decrease as $r$ rises. Consequently, the observed intensity shows a conspicuous peak resulting from the superposition of the photon ring, lensing ring, and direct emission since the emission source coincides with the region of intense gravitational lensing. However, in terms of spacetime geometry, the lensing ring and photon ring are still limited to a small area due to the demagnification factor. Nevertheless, for \textbf{Model 2}, the observation is still dominated by direct emission. The lensing ring still contributes only a small amount to the total flux, as can be seen from the right-hand column, although direct emission continues to be the dominant source.

As can be seen in the left column of the third row in Figs.\ref{fig:case1} and \ref{fig:case2}, when using \textbf{Model 3}, the emission extends inward, close to the event horizon $r_h$. Driven by the background metric, the range of the impact parameter $b$ that correspond to the lensing ring and photon ring are covered by the emission region specified in \textbf{Model 3}. Starting from just outside the event horizon, the observed intensity increases gradually, then rises sharply owing to the accumulation of the light rays, through the lensing ring region, and peaks within the photon ring region. The contribution from the lensing ring emission then causes the measured intensity to reach a lower peak. The observed intensity then starts to gradually decrease. Compared to the previous two models, the emission profile of \textbf{Model 3} ensures that the lensing ring's contribution to the total observed intensity is more noticeable in this case. Additionally, a brilliant ring is visible in the optical appearance, resulting from the geometric concentration of flux from the photon and lensing rings overlaid on the direct emission background.

The parameters of the Hernquist DM halo have a direct and significant impact on the observational features, as can be seen by a comparison of Figs.\ref{fig:case1} and \ref{fig:case2}. Geometrically, the photon and lensing ring radii both enlarge with increasing parameter since the crucial impact parameter $b_p$ has expanded. Radiatively, however, their contribution to the total flux diminishes for the emission models considered here. Furthermore, for \textbf{Model 3}, the diameter of the bright ring in the image appears to increase slightly related to the geometric expansion of the photon capture region. As the Hernquist DM halo parameter grows, the maximum measured intensity increases for all three emission scenarios ($2.12\%$ for \textbf{Model 1}, $6.22\%$ for \textbf{Model 2} and $5.31\%$ for \textbf{Model 3}) , whose behaviors are specific to the thin-disk configuration.


\section{Shadows with spherical accretions}
\label{sec4}
Having analyzed thin disk accretion, we now turn our attention to spherical accretion flows, which provide alternative astrophysical scenarios. We will investigate both static and infalling accretion models to determine how the Hernquist DM halo impacts the corresponding observational signatures. \revision{Unlike the thin disk case, for spherically symmetric accretion the shadow boundary is determined by the critical curve, and the classical notion of the black hole shadow applies directly~\cite{narayan2019shadow,falcke2000viewing}.}

\subsection{The static spherical accretion}
\begin{figure}[t]
\centering
\includegraphics[width=0.8\textwidth]{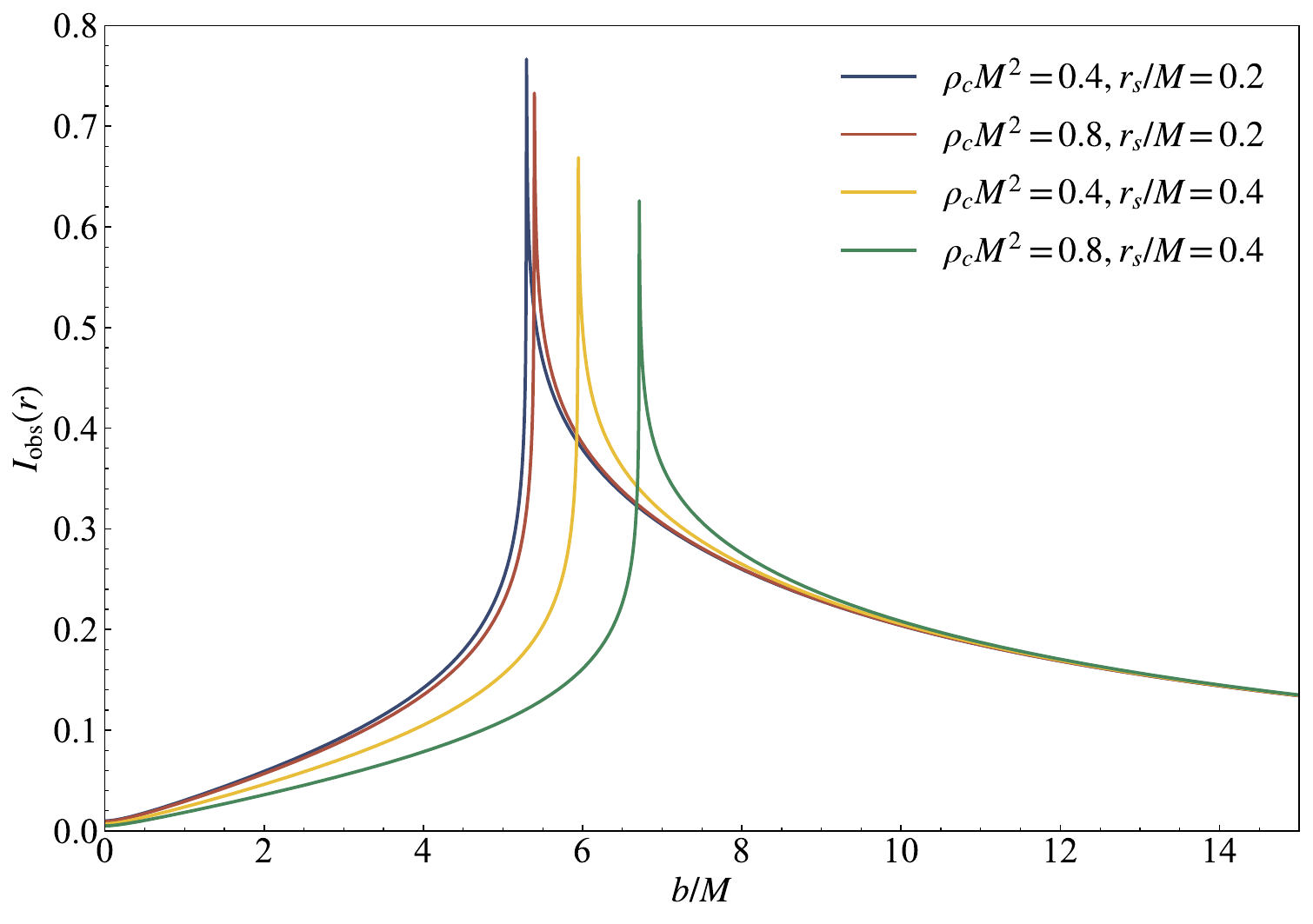}
\caption{The specific intensity $I_{\text{obs}}$ of a \textbf{static} spherical accretion as observed by a distant observer. The photon ring, where light rays circle the black hole several times before escaping, is represented by the sharp peak. The peak intensity reduces when the Hernquist DM parameters $\rho_c$ and $r_s$ increase. Here we fix $M=1$.}
\label{fig:static_1}
\end{figure}

\begin{figure}[t]
\centering
\includegraphics[width=\textwidth]{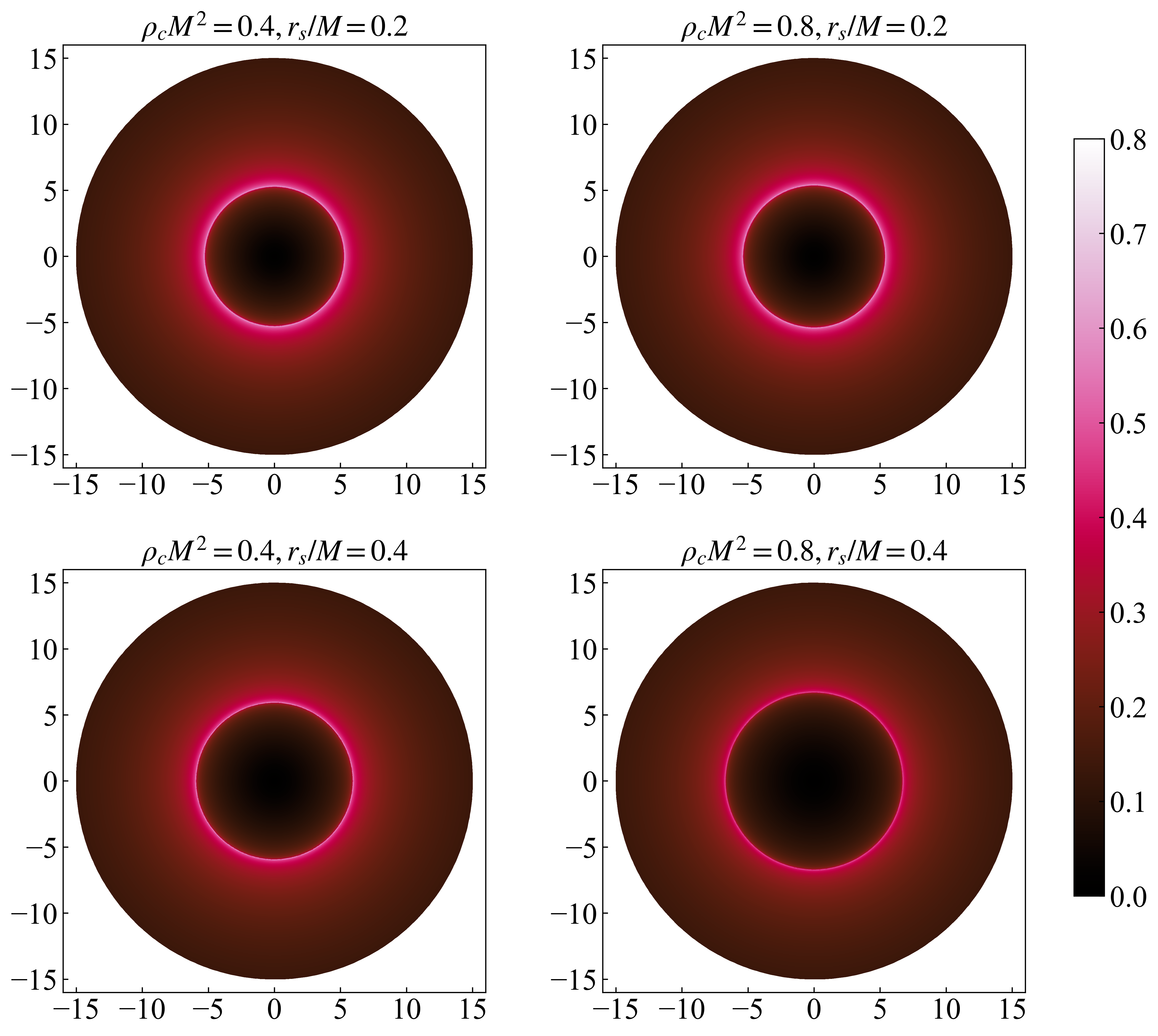}
\caption{The images of Schwarzschild-Hernquist black holes' shadow with \textbf{static} spherical accretion as observed by a distant observer. To enable direct quantitative comparison, all panels utilize a unified intensity color bar. The bright photon ring, which represents the peak of the intensity profile, surrounds the center of the dark area, which is the black hole shadow. As the halo parameters $\rho_c$ and $r_s$ increase, the shadow size expands, while the overall image brightness significantly decreases, as shown by the dimmer colors in the high-density panels of the Hernquist DM.}
\label{fig:static_2}
\end{figure}

We first study the shadow and photon ring in the scenario of a static spherical accretion surrounding a Schwarzschild-Hernquist black hole. The specific intensity observed at infinity is given by~\cite{jaroszynski1997optics,bambi2013can,fathi2023observational}
\begin{align}
\label{intensity}
I_{\text{obs}}=\int_\gamma \mathcal{R}^3\mathcal{J}\left(\nu_{\text{emi}}\right)\mathrm{d}\ell_{\text{prop}},
\end{align}
with
\begin{align}
\mathcal{J}\left(\nu_{\text{emi}}\right)\propto\dfrac{\delta\left(\nu_{\text{emi}}-\nu_{\text{fix}}\right)}{r^2},
\end{align}
over the trajectory of emitted photons $\gamma$. Here, $\mathrm{d}\ell_{\text{prop}}$ is the infinitesimal proper length, $\mathcal{J}\left(\nu_{\text{emi}}\right)$ is the emissivity per unit volume in the rest frame of the accreting material, $\nu_{\text{emi}}$ is the emitted photon frequency, and $\nu_{\text{fix}}$ is the fixed, monochromatic emission frequency. In the Schwarzschild-Hernquist spacetime, the redshift factor is
\begin{align}
\mathcal{R}=\dfrac{\nu_{\text{obs}}}{\nu_{\text{emi}}}=\sqrt{f(r)},
\end{align}
and the infinitesimal proper length is
\begin{align}
\label{dlprop}
\mathrm{d}\ell_{\text{prop}}=\sqrt{\dfrac{1}{f(r)}+r^2\left(\dfrac{\mathrm{d}\varphi}{\mathrm{d}r}\right)^2}\mathrm{d}r.
\end{align}
Combining Eqs.\eqref{intensity}-\eqref{dlprop}, the observed intensity for a static observer at infinity can be expressed as
\begin{align}
\label{intensity2}
I_{\text{obs}}=\int_{\gamma}\dfrac{f(r)^{3/2}}{r^2}\sqrt{\dfrac{1}{f(r)}+r^2\left(\dfrac{\mathrm{d}\varphi}{\mathrm{d}r}\right)^2}\mathrm{d}r.
\end{align}
The integrand in Eq.\eqref{intensity2} represents the intensity contribution as a function of impact parameters $b$, which is plotted in Fig.\ref{fig:static_1}. Fig.\ref{fig:static_2} displays the shadow and photon sphere of the Schwarzschild-Hernquist black hole by plotting the measured brightness versus $b$.

The resulting observed intensity $I_{\text{obs}}(b)$ first increases with the impact parameter $b$, peaks near the critical value $b\sim b_p$, and then gradually decreases for larger $b$. The observational intensity asymptotically approaches zero as $b$ approaches infinity. Furthermore, we find that the measured intensity is strongly influenced by changes in the Hernquist DM halo. As seen in Fig.\ref{fig:static_2}, the observed intensity's peak value decreases as $\rho_c$ and $r_s$ are increased. As shown in the images of Fig.\ref{fig:static_2}, the overall appearance becomes darker for larger values of $\rho_c$ and $r_s$. This brightness suppression is a radiative effect that is mostly caused by the dark matter halo's deepening potential well's higher gravitational redshift. The image for the case $\rho_cM^2=0.4$ and $r_s/M=0.2$ is significantly brighter than the image for $\rho_cM^2=0.8$ and $r_s/M=0.4$. Interestingly, unlike this radiative dimming, although the image becomes darker, the photon ring, which corresponds to the peak in $I_{\text{obs}}(b)$, exhibits a clear geometric expansion to larger values of $b$ as $\rho_c$ and $r_s$ increase, consistent with the increase in $b_p$ reported in Tab.\ref{tab:rh+rp+bp}. Keep in mind that the central intensity is not zero, but it is extremely modest, close to $b\sim0$. Due to the small but non-zero probability of radiation originating from very close to the black hole to escape, the region inside the photon ring in the images of Fig.\ref{fig:static_2} is not completely dark but exhibits a faint glow, brightest near the location of the photon ring itself.


\subsection{The infalling spherical accretion}
\begin{figure}[t]
\centering
\includegraphics[width=0.8\textwidth]{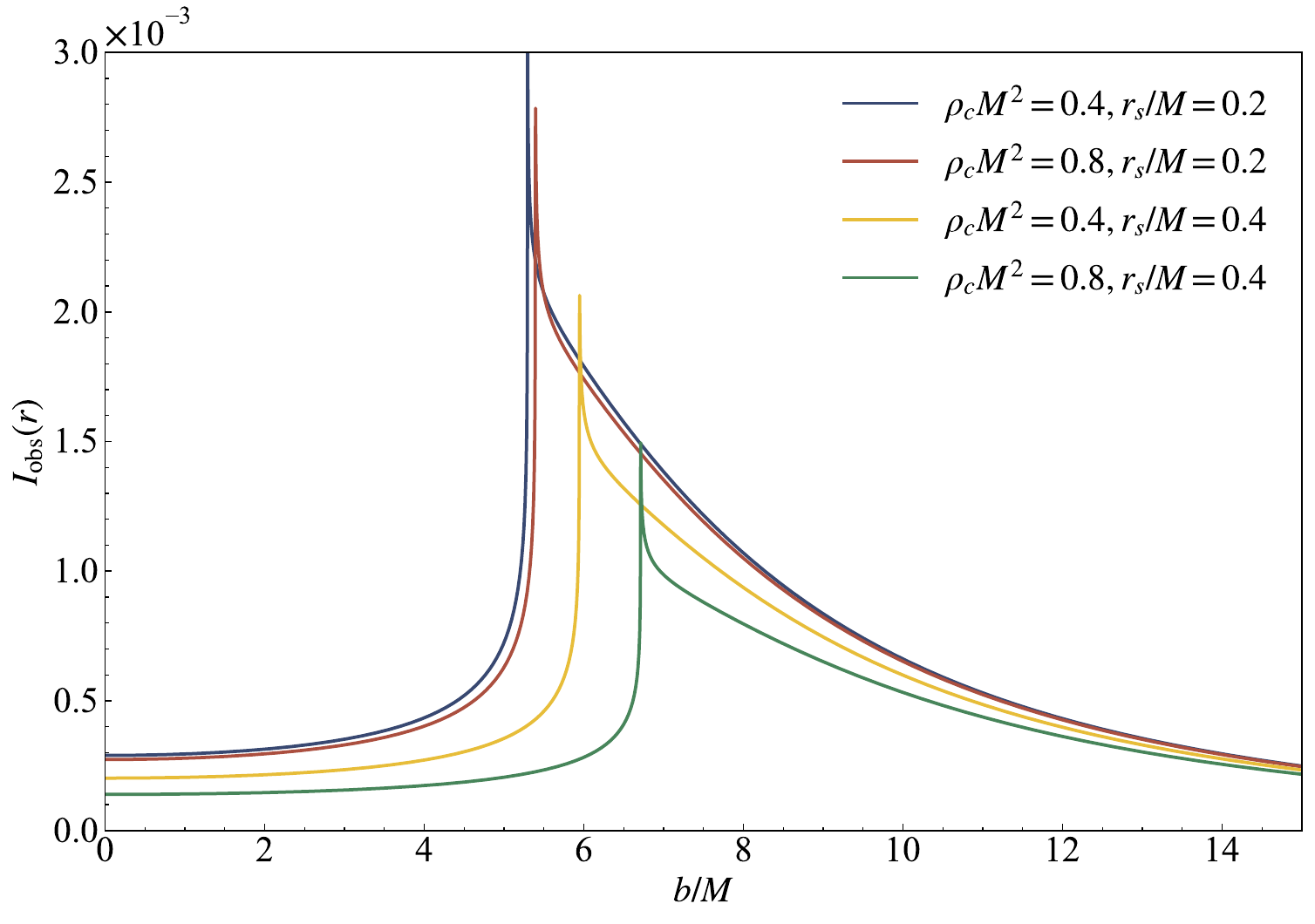}
\caption{The specific intensity $I_{\text{obs}}$ of an \textbf{infalling} spherical accretion as observed by a distant observer. The photon ring is represented by the sharp peak, similar to the static case,. The intensity is suppressed due to the Doppler de-boosting effect of the infalling matter. As the Hernquist DM parameters $\rho_c$ and $r_s$ increase, the peak intensity decreases significantly, while the position of the peak shifts to larger impact parameters $b$.}
\label{fig:infalling_1}
\end{figure}

We next investigate the case of infalling spherical accretion around Schwarzschild-Hernquist black holes. The general formula for the observed intensity, Eq.\eqref{intensity}, remains valid for infalling accretion. However, the redshift factor $\mathcal{R}$ must be modified to account for the velocity of the infalling matter. For infalling accretion, the redshift factor is given by~\cite{li2021observational,li2021shadows,fathi2023observational,bambi2013can}
\begin{align}
\label{redshift}
\mathcal{R}=\dfrac{\kappa_{\mu}u^{\mu}_{\text{obs}}}{\kappa_{\nu}u^{\nu}_{\text{emi}}},
\end{align}
The four-velocities for a distant static observer and for the infalling accreting matter are
\begin{align}
\bm{u}_{\text{obs}}=(1,0,0,0)
\end{align}
and
\begin{align}
\bm{u}_{\text{emi}}=\left(\dfrac{1}{f(r)},-\sqrt{1-f(r)},0,0\right),
\end{align}
respectively. In Eq.\eqref{redshift}, the four-velocities of photons $\bm{\kappa}$ released from the accretion disk is represented by the co-vector, which has the same definition as in Eqs.\eqref{dot_t}-\eqref{dot_r}. For purely radial accretion, we need to compute the ratio $\kappa_r/\kappa_t$. Using the photon's equations of motion, this ratio is found to be~\cite{bambi2013can}
\begin{align}
\dfrac{\kappa_r}{\kappa_t}=\pm\dfrac{1}{f(r)}\sqrt{1-\dfrac{b^2}{r^2}f(r)},
\end{align}
where the $\pm$ symbol indicates whether the photon is moving closer to or further away from the Schwarzschild-Hernquist black hole. Substituting these results into the expression for $\mathcal{R}$, the redshift factor becomes
\begin{align}
\mathcal{R}&=\left(u^{t}_{\text{emi}}+\dfrac{\kappa_r}{\kappa_t}u^{r}_{\text{emi}}\right)^{-1}\notag\\
&=\left[\dfrac{1}{f(r)}\pm\sqrt{\left(\dfrac{1}{f(r)}-1\right)\left(\dfrac{1}{f(r)}-\dfrac{b^2}{r^2}\right)}\right]^{-1}.
\end{align}

\begin{figure}[t]
\centering
\includegraphics[width=\textwidth]{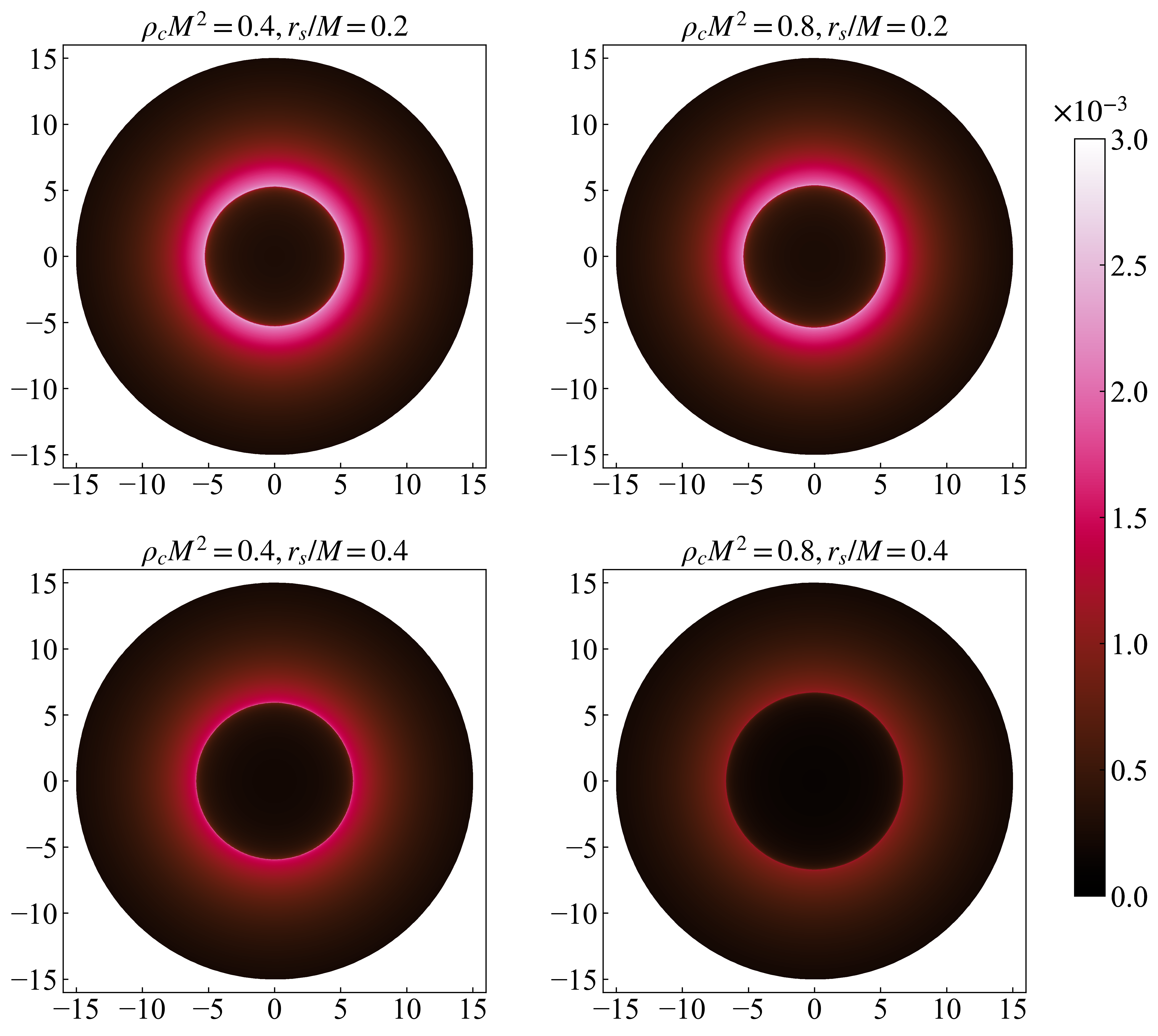}
\caption{The images of Schwarzschild-Hernquist black holes' shadow with \textbf{infalling} spherical accretion as observed by a distant observer. To enable direct quantitative comparison, all panels utilize a unified intensity color bar. The bright photon ring, which expands geometrically with increasing Hernquist DM parameters, outlines the shadow. Different from the \textbf{static} case, the central region exhibits pronounced darkness due to the Doppler de-boosting effect of the infalling matter.}
\label{fig:infalling_2}
\end{figure}

Similarly, we maintain the assumption that the emission is monochromatic. Thus, for the infalling accretion, the observed intensity takes the form
\begin{align}
\label{intensity3}
I_{\text{obs}}\propto\int_{\gamma}\dfrac{\mathcal{R}^3}{r^2}\dfrac{\kappa_t}{|\kappa_r|}\mathrm{d}r.
\end{align}
The observed intensity profile $I_{\text{obs}}$ and the corresponding images for infalling accretion, computed using Eq.\eqref{intensity3}, are shown in Figs.\ref{fig:infalling_1} and \ref{fig:infalling_2}, respectively.

As shown in Fig.\ref{fig:infalling_1}, and similar to the static case, the observed intensity $I_{\text{obs}}$ for infalling accretion peaks near the critical impact parameter $b_p$ for all Hernquist DM parameter. The intensity increases for $b<b_p$, rising sharply just before reaching the peak at $b\sim b_p$. For $b>b_p$, the intensity gradually decreases with increasing $b$, consistent with the behavior in the static model. The observed intensity at $\rho_cM^2=0.4$ and $r_s/M=0.2$ is substantially brighter than at $\rho_cM^2=0.8$ and $r_s/M=0.4$, as shown in Figs.\ref{fig:static_1} and \ref{fig:infalling_1}. A direct comparison between the images in Fig.\ref{fig:static_2} and Fig.\ref{fig:infalling_2} reveals that the central region is significantly darker for infalling accretion. When the parameters are changed from $\rho_cM^2=0.4$ and $r_s/M=0.2$ to $\rho_cM^2=0.8$ and $r_s/M=0.4$, the peak intensity decreases by $18.36\%$ for static accretion and by $50.50\%$ for infalling accretion. This suppression is a kinematic radiative effect caused by Doppler de-boosting, where the apparent intensity of radiation emitted by materials falling radially inward is reduced due to redshifting. Contrary to these radiative fluctuations, which are model-dependent, the geometric signature is robust. In both the static and infalling models, the radius of the photon ring increases with increasing Hernquist DM parameters, consistent with the findings in Tab.\ref{tab:rh+rp+bp}.


\section{Conclusion}
\label{conclusion}
In this work, we have conducted a thorough investigation of the optical appearance and shadow \revision{properties} of a Schwarzschild black hole embedded within a Hernquist DM halo. Our findings demonstrate that the presence of a Hernquist DM halo substantially alters the observable characteristics of the black hole \replace{shadow}{image} across a variety of accretion scenarios. In particular, the photon sphere expands significantly as $\rho_cM^2$ and $r_s$ increase, as seen in Figs.\ref{fig:case1}-\ref{fig:infalling_2}. Furthermore, we find that the core radius $r_s$ of the DM halo influences the growth of the photon sphere more significantly than the central density $\rho_c$. We use the Schwarzschild black hole with $b_p^{\text{Sch}}=3\sqrt{3}M$ as a reference case\cite{luminet1979image,gralla2019black}. For the parameter set $\rho_cM^2=0.4$ and $r_s=0.2$, the \replace{shadow diameter}{critical impact parameter $b_p$} increases relative to the Schwarzschild case by $\Delta\theta/\theta_{\text{Sch}}=\left(b_p^{\text{SH}}-b_p^{\text{Sch}}\right)/b_p^{\text{Sch}}=1.89\%$~\cite{perlick2022calculating}. Keeping $r_s=0.2$, the relative change rate $\Delta\theta/\theta_{\text{Sch}}$ rises to $3.78\%$ at $\rho_cM^2=0.8$. Notably, if we increase $r_s/M$ from $0.2$ to $0.4$ while keeping $\rho_cM^2=0.4$ fixed, the relative change climbs markedly to $14.40\%$. In the limiting case, the relative change rate $\Delta\theta/\theta_{\text{Sch}}$ reaches $29.15\%$ when $\rho_cM^2=0.8$ and $r_s=0.4$. However, as shown in Figs. \ref{fig:static_2} and \ref{fig:infalling_2}, the peak observed intensity is significantly diminished, resulting in a substantially dimmer overall image. The geometric expansion of the photon trajectory is controlled by the Hernquist DM halo's gravitational impact, although radiative transfer processes like gravitational redshift and Doppler de-boosting are the main causes of the intensity suppression. Our calculations indicate that while the presence of a Hernquist dark matter halo systematically increases the \replace{size of the shadow}{critical impact parameter} and the emission ring \revision{radius}, the dynamical state of the accretion flow (static versus infalling) has a comparable, if not more significant, influence on the absolute level and distribution of brightness. This suggests that, when using EHT observations to constrain the central dark matter distribution, it is imperative to have a prior assumption or an independent constraint on the physical state of the accretion flow.

The parameter space explored in this study requires particular attention to its physical significance, especially the core density $\rho_c M^2\in[0.4, 0.8]$ and core radius $r_s/M\in[0.2, 0.4]$. To ensure that spacetime near the photon sphere is not dominated by halo contribution, we could estimate the Herquist DM mass at the typical radius $r=5M$. Employing the Hernquist DM mass \eqref{DM_mass}, we find that the ratio of DM mass to BH mass, $(M_{\text{H}}/M_{\text{BH}})|_{r=5M}$, ranges from a negligible value $\sim 1.9\%$ for $r_s=0.2M$ and $\rho_c M^2=0.4$ to a significant yet sub-dominant $\sim 27.6\%$ for $r_s=0.4M$ and $\rho_c M^2=0.8$. This confirms that the geometry of the spacetime remains dominated by the central BH across the whole parameter space. We also compare our chosen parameters with Jha's suggested limitations with respect to astronomical observations~\cite{jha2025thermodynamics}. The parameters used in this study for the supermassive BH $\text{M87}^*$ fall within the $2\sigma$ upper limits obtained from EHT data. The same parameters, however, far surpass the more stringent limitations for $\text{SgrA}^*$. Thus, it is crucial to explicitly classify this model is a phenomenological toy model that investigates theoretical upper bounds. However, the analysis's findings still have definite ramifications for astronomical observations:
\begin{itemize}
\item Exclusion Region: The expected \replace{shadow deviation}{deviation of $b_p$} approaches $30\%$ in high-density areas. The probability of such high-density Hernquist DM regions in the central region of the Milky Way is essentially ruled out by this large signal, which is much larger than current observational errors of $\sim10\%$.
\item Feasibility Region: In contrast, the deviation becomes minor at $\sim1.89\%$, falling below observational error limits, when parameters adopt lower values. This suggests that the lower-density Hernquist DM distributions are still viable astrophysical candidates for explaining observed anomalies under the limitations of present technology.
\end{itemize}
Future facilities like ngEHT will improve precision. The implications of this work might place more stringent upper constraints on Hernquist DM density around the BH if no deviation is found with better accuracy. On the other hand, the low-density parameter space shown here may offer a convincing physical explanation for non-Kerr features if tiny deviations are found.

Importantly, our research offers significant insights into potential observable features in real astrophysical systems like $\text{M87}^*$, which are typically described by the Kerr metric~\cite{EHT1,EHT2,EHT3,EHT4,EHT5,EHT6,EHT7,EHT8,EHT9}, even though our model is based on a static, modified Schwarzschild background. The predicted variation in \replace{shadow}{the critical impact parameter} size, which manifests as a positive correlation between the photon ring diameter and Hernquist DM parameters, serves as a valuable theoretical framework for probing the DM distribution around supermassive black holes. The current EHT observation for $\text{M87}^*$ is approximately $42\pm3\,\mathrm{\mu as}$, whose relative uncertainty is about $7.14\%$~\cite{EHT1}, which is consistent with the Kerr vacuum solution. Our model provides a theoretical prediction that may be tested by current or next-generation EHT measurements, since this stands in stark contrast to the $\sim2\%$ to $\sim30\%$ rise anticipated by Hernquist DM halos. Consequently, when an anomalously large shadow detected in the future indicate the presence of a Hernquist DM halo, our current analysis suggests that the absence of significant shadow enlargement already places stringent constraints on the parameter space of viable Hernquist DM models.

\revision{It is instructive to compare our findings with related studies of dark matter effects on black hole imaging. Ref.~\cite{macedo2024optical} investigated the optical appearance of black holes surrounded by a Hernquist-type dark matter halo using the exact non-perturbative solution reported in Ref.~\cite{cardoso2022black}. Their analysis combined hot-spot astrometry and the Gralla-Lupsasca-Marrone emission models. For configurations saturating the EHT shadow-size constraint with compactness $\mathcal{C}\sim1/15$, they found that the photon ring locations shift outward by approximately $7\%$ and their widths increase by up to $\sim25\%$ compared to Schwarzschild case, while the relative luminosities of successive photon rings are only marginally affected. Their overall conclusion is that the optical appearance of the dark matter halo black hole closely mimics that of a Schwarzschild black hole under current observational precision. This is consistent with our finding that, for the lower-parameter configurations with $\rho_cM^2=0.4$ and $r_s/M=0.2$, the critical curve enlargement is merely $\sim2\%$, falling below the EHT detection threshold. More recently, Ref.~\cite{feng2026shadow} conducted a comprehensive study of the shadow and quasi-normal modes of the Schwarzschild-Hernquist black hole across a broad parameter space. They derived an upper bound on the halo compactness $\mathcal{C}\leq0.092$ from EHT data and obtained an analytical formula for the QNM frequency redshift which is valid for $\mathcal{C}\leq0.3$. This QNM redshift is driven by the same gravitational potential deepening that causes the shadow enlargement and brightness suppression reported in our work, and provides a complementary gravitational-wave channel for constraining the Hernquist DM halo parameters. Our work extends these previous studies by providing a systematic comparison across three distinct accretion scenarios, namely thin disk, static spherical, and infalling spherical accretion. This comparison reveals that while the geometric signature of the critical curve enlargement is robust across all models, the dynamical state of the accretion flow has a comparable, if not greater, influence on the brightness distribution. This model-dependent radiative behavior underscores the importance of incorporating realistic accretion physics when using black hole images to constrain dark matter distributions.}

Looking forward, our results highlight the potential of next-generation very-long-baseline interferometry (VLBI)~\cite{hobiger2009integer,schuh2012very,niell2018demonstration,nothnagel2018very} arrays to refine DM models through precise shadow imaging. Furthermore, they demonstrate a distinct, observable correlation between DM parameters. Future work incorporating black hole spin in Kerr-like spacetimes~\cite{ferrer2017dark,hou2018rotating} and more realistic, dynamic accretion flows will be essential to fully bridge the gap between theoretical predictions and observational evidence. Our results strongly support the use of black hole shadows as unique probes that can reveal the distribution and properties of DM halos, as well as elucidate new aspects of gravitational physics in strong-field regimes. This approach promises to shed new light on the nature of DM in galactic centers with increasing precision.


\vspace{1cm}
\noindent \textbf{Acknowledge}

This work is partly supported by the Shanghai Key Laboratory of
Astrophysics 18DZ2271600.

\newpage
\bibliography{reference}

\end{document}